\documentclass[%
reprint,
superscriptaddress,
groupedaddress,
amsmath,amssymb,
aps,
prx
]{revtex4-2}

\usepackage{dcolumn}% Align table columns on decimal point
\usepackage{bm}% bold math
\usepackage[colorlinks=true, allcolors=blue]{hyperref}% add hypertext capabilities
\usepackage[mathlines]{lineno}% Enable numbering of text and display math
\usepackage{todonotes}
\usepackage{acronym}
\graphicspath{ {./figures/} }
\usepackage[LGRgreek]{mathastext}

\begin{document}

%\preprint{APS/123-QED}

%\title{Universality of Polariton Chern bands in 2D PhC }
\title{Polariton Chern Bands in 2D Photonic Crystals Beyond Dirac Cones}

\author{Xin Xie$^{1,2}$}\email{xiexin@umich.edu}
\author{Kai Sun$^{1}$}\email{sunkai@umich.edu}
\author{Hui Deng$^{1}$}\email{dengh@umich.edu}%
\affiliation{$^{1}$Department of Physics, University of Michigan, Ann Arbor, Michigan 48109, United States\\}%Lines break automatically or can be forced with \\
\affiliation{$^{2}$Michigan Institute for Data Science, University of Michigan, Ann Arbor, Michigan 48109, United States\\}%Lines break automatically or can be forced with \\
%\affiliation{$^{3}$Applied Physics Program, University of Michigan, Ann Arbor, Michigan 48109, United States\\}
%\affiliation{$^{4}$Department of Electrical Engineering and Computer Science, University of Michigan, Ann Arbor, Michigan 48109, United States\\}


\begin{abstract}
Polaritons, formed by strong light-matter interactions, open new avenues for studying topological phases, where the spatial and time symmetries can be controlled via the light and matter components, respectively. 
However, most research on topological polaritons has been confined to hexagonal photonic lattices featuring Dirac cones at large wavenumbers. This restricts key topological properties and device performance, including sub-meV gap sizes that hinder further experimental investigations and future applications of polariton Chern insulator systems. 
In this study, we move beyond the traditional Dirac cone framework and introduce two alternative band structures in photonic crystals (PhCs) as promising platforms for realizing polariton Chern bands: bands with symmetry-protected bound states in the continuum (BICs) and bands with symmetry-protected degeneracies at the $\Gamma$ points. 
These band structures are prevalent in various PhC lattices and have features crucial for experimental studies. 
We show examples of higher Chern number bands, more uniform Berry curvature distributions, and an experimentally feasible system capable of achieving a large topological gap. 
Our findings show the broad applicability of polariton Chern bands in 2D PhCs, provide design principles for enhancing the functionality and performance of topological photonic devices, opening up exciting possibilities for better understanding and using topological physics.
\end{abstract}

\maketitle

%\tableofcontents

\section{\label{sec:level1}Introduction}
The topological classification of band structures has revolutionized our understanding of novel quantum states resulting from special symmetry properties. 
Photonic systems have formed a counterpart to condensed matter systems for exploring topological phenomena, where different geometries can be implemented using periodic photonic structures~\cite{Lu_Topological_2014,Khanikaev_Twodimensional_2017,Ozawa_Topological_2019,Smirnova_Nonlinear_2020,Kim_Recent_2020}. 
Celebrated examples include photonic analogues of quantum spin Hall and quantum valley Hall effects in systems with crystalline symmetry but preserved time-reversal (TR) symmetry~\cite{Umucalilar_Artificial_2011,Hafezi_Robust_2011,Hafezi_Imaging_2013,Wu_Scheme_2015,Dong_Valley_2017,Shalaev_Robust_2019,Liu_Generation_2020b,Li_Experimental_2021a}.
%\xin{quantum spin Hall don't require broken inversion symmetry. So I replace it with crystalline symmetry.}
%Umucalilar_Artificial_2011,Hafezi_Robust_2011:  the gauge field in these systems does not break TR symmetry, belonging to quantum spin Hall effect. 
These systems support gap opening and edge states, although topological protections are limited and only at carefully designed boundaries due to the absence of full-band topology~\cite{Arregui_Quantifying_2021,Rosiek_Observation_2023,Khanikaev_Topological_2024}. 
%\hui{*Need improvement.}

To obtain Chern bands with fully topologically protected edge states, breaking the TR symmetry is required~\cite{Haldane_Possible_2008,Raghu_Analogs_2008}. The first demonstrations include using gyromagnetic photonic crystals (PhCs) in the microwave regime, which exhibited unidirectional transport of microwaves with no back reflection even amidst substantial disorders
~\cite{Wang_Observation_2009}. Extending Chern bands to optical frequencies would greatly broaden the technological potentials of such systems, but magnetic permittivity is typically very weak at optical frequencies. 
Solutions to circumvent the difficulty of breaking TR symmetry in optical systems include introducing additional spatial dimensions to simulate the time dimension~\cite{Rechtsman_Photonic_2013,Lustig_Photonic_2019}, using time-periodic driving fields~\cite{Lindner_Floquet_2011,Fang_Realizing_2012, Maczewsky_Observation_2017}, and engineering gain and dissipation~\cite{Zhao_NonHermitian_2019, Xiao_NonHermitian_2020, Dai_NonHermitian_2024}.
%To address this, alternative designs have been developed, including time-periodic systems~\cite{Fang_Realizing_2012, Rechtsman_Photonic_2013, Maczewsky_Observation_2017}, non-Hermitian systems~\cite{Zhao_NonHermitian_2019, Xiao_NonHermitian_2020, Dai_NonHermitian_2024}, 
An alternative route to create true Chern bands is to hybridize cavity photon modes with exciton modes to form polaritons~\cite{weisbuch_observation_1992} and utilize excitons to break the TR symmetry~\cite{Karzig_Topological_2015}.

In the hybridized polariton system, TR symmetry breaking can be achieved by lifting the degeneracy of the spin states of excitons, while spatial symmetries can be engineered via the photon modes to achieve phase winding and nontrivial band topology. 
Following the initial concept of topological polaritons~\cite{Karzig_Topological_2015}, theoretical proposals have emerged for 2D planner cavities~\cite{Gianfrate_Measurement_2020,Polimeno_Tuning_2021}, coupled micropillars~\cite{Nalitov_Polariton_2015,Bardyn_Topological_2015}, and 2D PhCs~\cite{Yi_Topological_2016,He_Polaritonic_2023}. 
Among these, the hexagonal micro-pillar lattice~\cite{Nalitov_Polariton_2015} corresponds to an experimentally viable system, where a complete topological gap can be opened by a magnetic field at Dirac points, leading to the first, and only, experimental demonstration of a polariton Chern insulator~\cite{Klembt_Excitonpolariton_2018a}. 
%Among these, complete topological gaps have been predicted for photonic graphene analogues hosting Dirac cones at $K/K'$
%~\cite{Nalitov_Polariton_2015,Bardyn_Topological_2015,Yi_Topological_2016,He_Polaritonic_2023},
%with the seminal experimental demonstration using coupled micropillar lattices~\cite{Klembt_Excitonpolariton_2018a}. 
%In the presence of spin-orbit coupling, these Dirac cones in coupled micropillar lattice can transition into quadratic touching bands, leading to topological polariton bands with Chern number of $\pm2$~\cite{Nalitov_Polariton_2015,Bardyn_Topological_2015,Klembt_Excitonpolariton_2018a}.
However, the topological gap is limited to about 0.1~meV due to the relatively large size of the micropillars~\cite{Klembt_Excitonpolariton_2018a}. 
Further research and application of polariton Chern insulators call for a system with a larger topological gap, which remains an outstanding challenge. 
2D PhCs have a larger Brillouin zone and could allow larger gaps at Dirac points, but these Dirac points are below the light line, compromising experimental feasibility~\cite{Yi_Topological_2016,He_Polaritonic_2023}.
More generally, the narrow focus on Dirac cones in hexagonal lattices limits the optimization of topological properties and the exploration of a wider range of topological phenomena. 
%These limitations of systems based on Dirac cones pose challenges in device performance and experimental accessibility.
%However, Dirac cones typically appear in high-momentum regions of specific PhCs, such as honeycomb or triangular lattice with $C_6$ symmetry. %In particular, the Dirac cones are beyond the light cone in 2D slab, posing great chanllenge in experimental accessiblity. 
%But the focus on Dirac cones limits the design space and challenges the optimization of performance and functionality of topological photonic devices. Hence, expanding the design landscape beyond Dirac cones and exploring diverse photonic bands is crucial for realizing new possibilities and improving device performance.
A larger design space of photonic bands will enrich polaritonic topological phenomena and facilitate technological applications. 

%For instance, coupled micro-pillars embedding quantum wells have demonstrated a 0.1 meV topological gap under a 5 T magnetic field~\cite{Klembt_Excitonpolariton_2018a}. A larger gap of 4 meV has been theoretically predicted recently by strongly coupling valley excitons with a substantial 40 meV splitting in monolayer TMDs to the Dirac cone around K point in 2D PhC~\cite{He_Polaritonic_2023}. 
%typically through the application of a magnetic field. Additionally, the valley-selective optical Stark effect can induce considerable splitting between valley excitons in transition metal dichalcogenides (TMDs) with circularly polarized pumps~\cite{Kim_Ultrafast_2014,Sie_Valleyselective_2015,Cunningham_Resonant_2019,Yong_Valleydependent_2019,Lyons_Giant_2022}, providing a more efficient approach to breaking TR symmetry. 

%In this work, we explore the formation of Chern bands by examining how strong coupling between excitons in 2D semiconductors and various photonic bands in 2D PhCs can lead to topologically nontrivial states. 
%We begin by introducing the conventional scenario involving Dirac cones and then extend our discussion to other prominent bands found in 2D PhCs. Specifically, we investigate: (1) gapped quadratic touching bands near the $\Gamma$ point and (2) polarization vortices around bound states in the continuum (BICs), as shown in Fig.~\ref{f1}. 
In this work, going beyond Dirac cones, we demonstrate two general types of PhC band structures -- symmetry-protected bound states in the continuum (BICs) circled by polarization vortices and symmetry-protected quadratic touching bands -- 
%\textcolor{red}{%
as effective platforms for realizing polariton Chern insulators in diverse 2D PhC lattices with C3, C4, and C6 symmetries. The topological gap and edge states are located near the $\Gamma$ point, with distinct ring-shaped Berry curvature distributions, which facilitates experimental accessibility and Floquet band engineering. 
%We further examine the dependence of topological properties on design parameters and demonstrate how to optimize PhCs for larger topological gaps.
We show examples of experimentally realizable PhC-polariton systems that respectively feature higher Chern numbers and a large 12~meV topological gap for high-temperature Chern insulator. These results provide a general framework for photonic Chern insulator systems and highlight the potential of 2D PhCs for achieving polariton Chern bands with new phenomena and improved device performance.

%\section{Results}
\section{General principle}
The general principle underlying the formation of polariton Chern bands can be captured by a two-component spinor model, where the winding coupling along with TR symmetry breaking leads to nontrivial Chern numbers. The Hamiltonian of the model is expressed as:
\begin{equation}
H_{\mathbf{k}} =
\begin{pmatrix}
-\Delta\omega/2 & g_{k} e^{-i\Phi_{\mathbf{k}}} \\
g_{k}^{*} e^{i\Phi_{\mathbf{k}}} & \Delta\omega/2
\end{pmatrix}.
\label{eq:H}
\end{equation}
Here, \(\Delta\omega\) denotes the energy difference between the two states, \(g_{k}\) is the amplitude of the coupling strength (with \(g_{k} \rightarrow 0\) as \(k\rightarrow 0\)), and \(\Phi_{\mathbf{k}}\) is phase of coupling. The distinct symmetry of two basis results in the winding coupling \(\Phi_{\mathbf{k}}=\textit{m}\phi_{\mathbf{k}}\), where \(\textit{m}\) is a nonzero integer, $\phi_{\mathbf{k}}$ is the azimuthal angle of the momentum \(\mathbf{k}\). 

\begin{figure}
\centering
\includegraphics[width=1\linewidth]{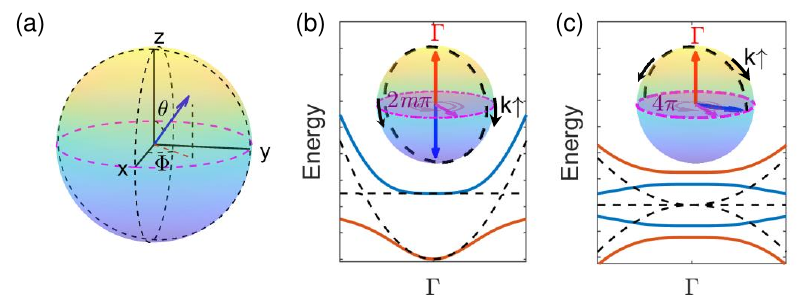}
\caption{\label{f1} (a) Spinor character of eigenstates parameterized on a Bloch sphere by the polar angle $\theta$ and azimuthal angle $\Phi$. (b-c) Polariton Chern bands (red and blue lines) realized by strong coupling of single exciton band with two types of PhC bands (black dashed lines): (b) bands with polarization vortex around symmetry-protected BIC and (c) bands with symmetry-protected quadratic touching degeneracy at $\Gamma$ points. Insets illustrate the evolution of eigenstates with momentum $\mathbf{k}$ from $\Gamma$ (red arrows) to high k (blue arrows) on Bloch sphere, displaying a full or a half loop (black dashed lines) accompanied with winding of $2\textit{m}\pi$ or $4\pi$ (magenta dashed circles), respectively.  
%\hui{Need to add exciton line to (c), and may need to update arrows etc. to show more clearly how the states move on the spheres. y-label of (b) is partially blocked.}\xin{modified accordingly}
}
\end{figure}

%To reveal the nontrivial topology, 
The Hamiltonian $H_k$ has eigenvalues: \(\pm \sqrt{g_k^2+\Delta\omega^2/4}\). The corresponding eigenstates are:
\begin{equation}
\label{eq:S}
\left| u_{+}(\mathbf{k}) \right\rangle = \begin{pmatrix}
\cos\left(\frac{\theta_{\mathbf{k}}}{2}\right) \\
e^{i\textit{m}\phi_{\mathbf{k}}} \sin\left(\frac{\theta_{\mathbf{k}}}{2}\right)
\end{pmatrix};
\end{equation}
\begin{equation}
\left| u_{-}(\mathbf{k}) \right\rangle = \begin{pmatrix}
\sin\left(\frac{\theta_{\mathbf{k}}}{2}\right) \\
-e^{-i\textit{m}\phi_{\mathbf{k}}} \cos\left(\frac{\theta_{\mathbf{k}}}{2}\right)
\end{pmatrix},
\end{equation} 
which can be represented by a spinor on the Bloch sphere with the polar angle $\theta=\theta_{\mathbf{k}}$ and azimuthal angle $\Phi=\textit{m}\phi_{\mathbf{k}}$, as shown in Fig. \ref{f1}(a). Here, the polar angle $\theta_{\mathbf{k}}$ is given by:
\begin{equation}
\theta_{\mathbf{k}}=arctan(\frac{2g_k}{\Delta\omega}).
\end{equation}
Broken TR symmetry leads to a non-zero $\Delta\omega$, lifting the degeneracy of the two states. The winding term $e^{i\textit{m}\phi_{\mathbf{k}}}$ causes the spinors to trace a path around the Bloch sphere $\textit{m}$ times. Depending on whether the path completes a full or half loop, the resulting Chern number is $\textit{m}$ or $\textit{m}/2$, as exemplified by the insets of Figs. \ref{f1}(b-c).

%\hui{Xin: perhaps we can focus on the quadratic touching and BIC and only mention Dirac cone as a reference. In this case, I wonder if we can replace Fig 1a by a Bloch sphere illustration of the pseudo-spin and winding. Ideally Fig 1b and 1c could help visually correlate the two cases with the winding in Fig 1a. (Right now, 1b and 1c are repeated in Fig 2 and Fig.6, which typically should be avoided in publications.) }
%\xin{revised accordingly}

%\hui{*Shall we use "class" or "type".}
Using PhC-polaritons, we can implement the Hamiltonian $H_k$ in two types of system, where we use PhCs to control spatial symmetry and excitons to break TR symmetry. 
The first type (Fig. \ref{f1}(b)) involves a non-degenerate PhC band with a symmetry-protected BIC at $\Gamma$. The BIC is surrounded by a polarization vortex, which introduces winding coupling between the photon and exciton bands. A topological gap is opened when the exciton band breaks TR symmetry.
The second type (Fig. \ref{f1}(c)) involves a pair of PhC bands, with symmetry-protected degeneracy points and winding coupling between the two PhC bands around the degeneracy points. Strong coupling with excitons allows breaking TR symmetry through the excitons and thereby opening a topological gap. Such degeneracy points include both the widely studied Dirac points at $K/K'$ in hexagonal lattices and the quadratic touching points at $\Gamma$ that we discuss in this work.  
%\xin{This part doesn't include the information related to the insets in Fig. \ref{f1}(b-c). Should we briefly introduce the difference of the evolution of eigenstates and the resulting Chern number for the two cases here?}
%\hui{Sure. Perhaps integrate with the discussions above about the two types.}
%\xin{add one sentence at the end of the following paragraph: as exemplified by the insets of Figs. \ref{f1}(b-c)}

Below we discuss the effective Hamiltonian and topological properties of each type, propose practical designs, and validate them through numerical simulations. %Our results predict a large gap of around 12 meV using quadratic touching bands, significantly enhancing performance. These findings highlight the potential of 2D PhCs to achieve polariton Chern bands and emphasize the advantages of exploring various photonic bands to enhance the performance and functionality of future photonic devices.

\section{Symmetry-protected BIC with a winding polarization vortex}
\begin{figure}
\centering
\includegraphics[width=1\linewidth]{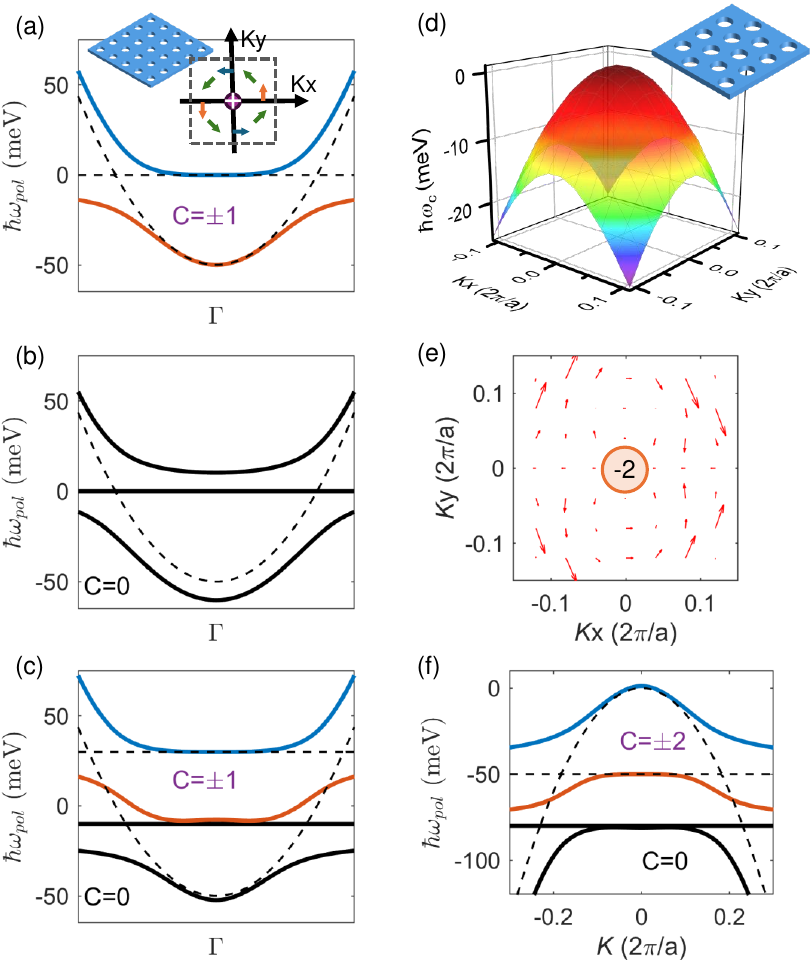}
\caption{\label{f6_sch_BIC} (a) Topological polariton formed by coupling the polarization vortex around a BIC to a single bright valley exciton. The inset shows the schematic of the polarization vortex around a symmetry-protected BIC at $\Gamma$ in PhCs with a square lattice. (b) Trivial polariton bands due to the existence of dark excitons. (c) Topological polariton in the case where only a bright exciton is dressed. (d) TE band with topological charge of $-2$ in PhC with triangular lattice (inset). (e) Distribution of polarization vectors of a polarization vortex with a topological charge of $-2$. (f) Topological polariton bands with Chern number of $\pm2$. The black dashed lines represent the uncoupled case, while the solid lines represent the polariton bands. The blue and red lines correspond to nontrivial bands with opposite Chern numbers, and the black solid lines correspond to trivial bands.}
\end{figure}

%\begin{figure}
%\centering
%\includegraphics[width=1\linewidth]{f7_BIC2.png}
%\caption{\label{f7} (a) TE band with a topological charge of $-2$ in PhC with triangular lattice (inset). (b) Distribution of polarization vectors displaying a polarization vortex with a topological charge of $-2$. (c) Topological polariton with Chern number of $\pm2$ formed by coupling polarization vortex around a BIC to a single valley exciton. (d) Topological polariton scenario where only "bright" excitons are dressed.  }
%\end{figure}

%Besides gapping the symmetry-protected band touching of two photonic bands, a second type of band structures can support polariton Chern bands, involving winding coupling between a single bright exciton band and a single photonic band with symmetry-protected BIC at $\Gamma$, as depicted in Fig. \ref{f1}(c). 

%The BIC is surrounded by a polarization vortex that winds $\textit{m}$ times around the BIC. Unlike the first type, which requires two photonic bands with symmetry-protected degeneracy, this type only requires a single photonic band. The resulting polariton bands evolve from more photon-like to more exciton-like, or vice versa, with increasing momentum. 
%The winding coupling induced by the polarization vortex causes the polariton bands to undergo a full wrapping of the Bloch sphere $\textit{m}$ times, leading to Chern numbers of $\pm\textit{m}$, as depicted in Fig. \ref{f1}(c).

Photonic BICs are modes that reside within the radiation continuum but are perfectly confined without radiating~\cite{vonNeumann_Ueber_1993,Hsu_Bound_2016,Kang_Applications_2023}. They commonly exist in 2D PhCs and can be categorized into symmetry-protected BICs and accidental BICs~\cite{Hsu_Observation_2013,Hsu_Bound_2016,Jin_Topologically_2019,Kang_Merging_2022,Kang_Applications_2023}. 
Symmetry protected BICs result from distinct symmetry classes between the BIC at high symmetry points and continuum modes~\cite{Hsu_Bound_2016,Kang_Applications_2023}.
They are surrounded by a polarization vortex that winds $\textit{m}$ times around the BIC, corresponding to a conserved and quantized topological charge $\textit{m}$~\cite{Zhen_Topological_2014,Zhang_Observation_2018,Yoda_Generation_2020a}.
%~\cite{Zhen_Topological_2014,Doeleman_Experimental_2018}.
% \hui{Add/check references if applicable.} \xin{checked}
An example is the $\Gamma$ point of a non-degenerate band in PhCs with $C_{4\nu}$ symmetry, as shown in Fig.~\ref{f6_sch_BIC}(a).  
There has been intense research on BICs, such as using them for vortex beams~\cite{Wang_Generating_2020,Huang_Ultrafast_2020}, polariton condensation~\cite{Ardizzone_Polariton_2022,Wu_Exciton_2024}, chirality~\cite{Liu_Circularly_2019,Chen_Observation_2023a}, lasing~\cite{Kodigala_Lasing_2017,Hwang_Ultralowthreshold_2021,Yu_Ultracoherent_2021,Ren_Lowthreshold_2022}, nonlinear optics~\cite{Liu_High_2019,Koshelev_Subwavelength_2020,Schiattarella_Directive_2024}, and sensing~\cite{Liu_Phase_2023}, etc. However, their potential for realizing quantum Hall phases has not been studied. 
Here, we show that strong coupling between symmetry-protected BICs with excitons can be used to form polariton Chern bands and, furthermore, allow high Chern numbers that are difficult to obtain in other configurations.

To analyze strong coupling with BIC modes, we first consider the simplified scenario where a single circularly-polarized exciton band is coupled to a photon band with a symmetry-protected BIC of charge $\textit{m}$. Other exciton modes are either shifted far away or saturated. 
The system can be described by the effective Hamiltonian:
\begin{equation}
    H_{\mathbf{k}}=
    \begin{bmatrix}
        \omega_{\mathbf{k}}^c & g_{k}e^{-i\Phi_{\mathbf{k}}} & \\ \\
        g_{k}^*e^{i\Phi_{\mathbf{k}}} & \omega_{K}^x \\        
    \end{bmatrix}.
\end{equation}
%\xin{this Hamiltonian here aims to explain the nontrivial topology induced by BIC. Focusing on a single exciton mode simplifies the analysis, making the topological features more comprehensible. However, if we include two split exciton modes, both would need to be bright, potentially causing confusion with the following scenario where only one bright exciton mode is shifted. }
Here, $\omega_{\mathbf{k}}^c$ and $\omega_{K}^x$ correspond to cavity and exciton dispersions, respectively. $g_k$ is the amplitude of the exciton-photon coupling strength; due to the symmetry mismatch between the bright exciton and the non-radiative BIC, $g_{\mathbf{k}}\rightarrow 0$ as $\mathbf{k}\rightarrow 0$. 
The polarization vortex surrounding the BIC gives rise to the winding phase in the coupling: $\Phi_{\mathbf{k}}=\textit{m} \phi_{\mathbf{k}}$, which corresponds to the polarization angle of the photon mode at ${\mathbf{k}}$. %, and $\textit{m}$ is the topological charge of the BIC. % for a BIC with a topological charge $\textit{m}$. 
%Due to the symmetry mismatch between the bright exciton and the non-radiative BIC, $g_{\mathbf{k}}\rightarrow 0$ as $\mathbf{k}\rightarrow 0$. 
The winding coupling results in polariton bands that wind in momentum space around the BIC, acquiring nontrivial topology. The lower and upper polariton bands transition from photon-like to exciton-like states with increasing or decreasing $|k|$, respectively. Consequently, the spinor traces a full wrap of the Bloch sphere $\textit{m}$ times, as shown in the inset of Fig. \ref{f1}(b), resulting in a Chern number of $\pm \textit{m}$ for the lower and upper polariton bands, as exemplified by Fig.~\ref{f6_sch_BIC}(a).

The above analysis ignores other exciton modes. In 2D PhC polariton systems, since the PhC BZ is a few orders of magnitude smaller than the exciton's, there are a large number of degenerate exciton modes within the photon Brillouin zone (BZ). Hence, %each photon mode couples to a bright exciton mode collectively while there exists a large number of other ``dark'' exciton modes that do not couple to this photon band. 
there exist dark excitons that couple to nonradiative modes, leading to hybridization and trivialization of polariton bands, as shown in Fig.~\ref{f6_sch_BIC}(b). However, Chern bands can be reconstructed by energetically separating bright excitons from dark ones. This is illustrated in Fig.~\ref{f6_sch_BIC}(c), where the bright exciton mode is shifted to a higher energy, such as by mode-selective optical Stark effect. The bright exciton mode can not hybridize with the BIC mode due to different symmetries, leading to a topological gap opening similar to the simplified case in Fig.~\ref{f6_sch_BIC}(a).
%\hui{Please double check or update the above description.}
%\xin{checked}

Notably, BIC systems are apt to realize polariton bands with higher Chern numbers. High-symmetry groups support high-order BICs with topological charges greater than one~\cite{Yoda_Generation_2020a,Kang_Merging_2022}. Figures \ref{f6_sch_BIC}(d-f) show a symmetry-protected BIC with $\textit{m}=-2$, at the $\Gamma$ point of a PhC with $C_{6\nu}$ symmetry. It gives rise to polariton bands with Chern numbers of $\pm2$, as illustrated in Fig. \ref{f6_sch_BIC}(f). 
%\textcolor{red}{
The realization of higher Chern number bands opens new opportunities to explore novel physical phenomena and potential applications~\cite{Skirlo_Multimode_2014,Lu_Topological_2016,Wu_Applications_2017}.
%\xin{These papers are electronic system of high Chern number bands.\cite{Trescher_Flat_2012,Yang_Topological_2012,Sterdyniak_Series_2013,Wu_Fractional_2015,Dong_Manybody_2023} I am not sure whether it is suitable to cite them here.}

\section{Gapping out symmetry-protected quadratic touching bands}
%\hui{Xin, can you try to rewrite this part. I'm thinking to start with a discussion that applies to both Dirac and quadritic touching, as they are described by similar physics and the same Hamiltonian (6), except different  g(k) and m. Then pointing out the difference, e.g. Dirac is linear dispersion (lowest order?) that only appears in pairs and thus generally in hexagon lattice at K and K'. Quadratic touching can appear at Gamma point and is common to all lattice symmetries (c3,4,6). Not sure if there is a reason to expect a larger gap for quadratic touching at Gamma, too, though that's our observation.
%Then focus on quadratic touching in A.1, optimization in A.2, practical realization of only the quadratic touching in A.3.}
%\xin{done}
\begin{figure}
\centering
\includegraphics[width=1\linewidth]{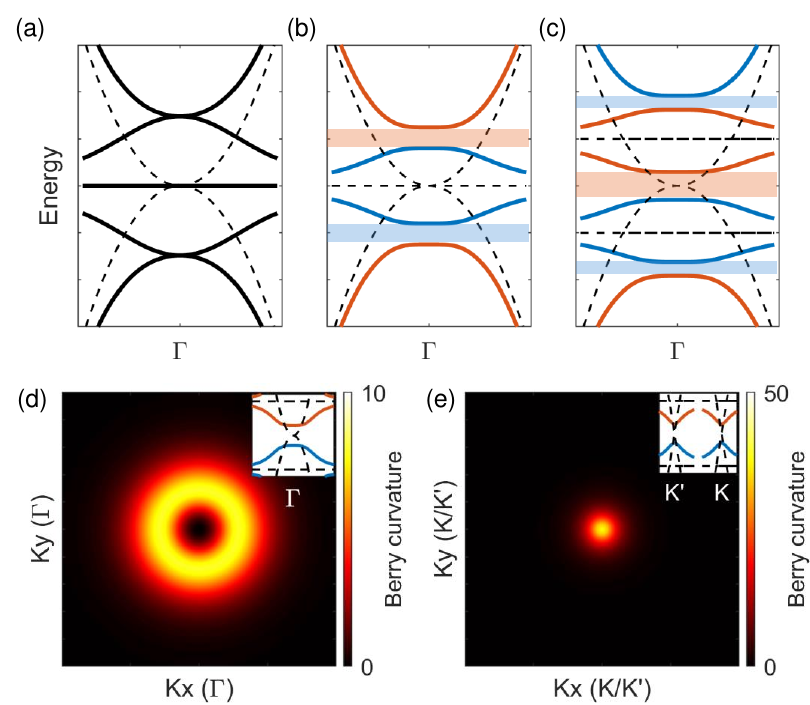}
\caption{\label{f_sch_quad} (a) Polariton band structure with TR symmetry, where the quadratic touching point is resonant to the two degenerate exciton bands. (b, c) TR symmetry-breaking cases with quadratic touching bands coupled to (b) a single exciton and (c) two split excitons. Black dashed lines show uncoupled states, and solid lines represent polariton bands. The red and blue areas highlight topological gaps. The colors encode Chern numbers: black for trivial, blue and red for nontrivial with opposite signs. (d, e) Berry curvature distributions for one of the topological polariton bands based on (d) quadratic touching bands and (e) Dirac cone around $\Gamma$ and $K/K'$ with the same $\mathbf{k}$ space scale. Insets show corresponding band dispersions. 
}
\end{figure}
Besides PhCs with BICs, polariton Chern bands can form in a second type of PhCs: PhCs with symmetry-protected degeneracies. The lowest orders are the Dirac points and quadratic-touching points. 

Quadratic-touching bands in 2D PhCs, similar to Dirac cones, have degeneracies at high-symmetry points that are protected by a combination of spatial and TR symmetries. Away from these high-symmetry points, the reduction in symmetry causes hybridization of the eigenmodes, resulting in a winding coupling of the form $g_k e^{i\textit{m}\phi_{\mathbf{k}}}$~\cite{Chong_Effective_2008,Mei_Firstprinciples_2012,Gao_Topological_2023a}. The introduction of exciton splitting breaks the TR symmetry, leading to a non-zero $\Delta\omega$ and gapping the degenerate bands. Together, the two effects lead to gapped bands with non-trivial Chern numbers. 

Although Dirac cones and quadratic-touching bands share this conceptual framework, they differ in the nature of symmetry classes and resulting winding coupling and Berry curvature distribution in the BZ. Dirac points are protected by inversion and TR symmetries, most commonly observed in pairs at $K$ and $K'$ of 
%at high-symmetry points with large momenta in PhCs with inversion symmetry. A well-studied example in
hexagonal lattices with $C_6$ symmetry~\cite{Ochiai_Photonic_2009,Collins_Integrated_2016}. The winding coupling increases linearly with $k$ as $\nu ke^{-i\phi_{\mathbf{k}}}$ with a winding number $\textit{m} = 1$, leading to a linear dispersion of the coupled modes with Berry flux of $\pm \pi$ around each Dirac point. Breaking TR symmetry results in an identical Berry flux contribution from each cone, yielding a nontrivial Chern number of $\pm 1$.

In contrast, symmetry-protected quadratic touching bands are commonly found in 2D PhCs with a variety of $C_{n>2}$ symmetries that belong to the $E$ representation of the point group, including $C_3$, $C_4$, and $C_6$. They can be described as rotational eigenmodes of the corresponding symmetries with eigenvalues of $\textit{e}^{\pm i2\textit{l}\pi/\textit{n}}$, where $\textit{l}$ is related to the distinct irreducible representations of their point group. They are degenerate at $\Gamma$ and have non-zero winding coupling that increases quadratically with $k$ as $\nu k^2e^{-i2\phi_{\mathbf{k}}}$ with a winding number $\textit{m} = 2$, leading to quratic dispersions of the coupled modes with Berry flux of $\pm 2\pi$ around $\Gamma$. Breaking TR symmetry results in
a half wrapping of the Bloch sphere $\textit{m}$ times, resulting in non-trivial Chern numbers of $\pm\textit{m}/2=\pm1$ for the gapped bands, as illustrated in Fig. \ref{f1}(c). 

The effective Hamiltonian of the PhC-polariton system can be written in a six-component basis $\Phi=(|P^+>,|\uparrow^+>,|\downarrow^+>,|P^->,|\uparrow^->,|\downarrow^->)$, 
%$\Phi=(|P^+>,|K^+>,|K'^+>,|P^->,|K^->,|K'^->)$
where $|P^{\pm}>$ are the two degenerate eigenmodes of the rotation operator of the PhC, $|\uparrow^{\pm}>$ and  $|\downarrow^{\pm}>$ are the collective excitonic modes with opposite spins and couple to $|P^{\pm}>$, respectively. 
%We derive the effective Hamiltonian of a PhC-polariton system with quadratic touching bands to analyze its topology. Since the PhC BZ is a few orders of magnitude smaller than the exciton's, there are a large number of degenerate exciton modes with the photon BZ. Hence, each photon band couples to a bright exciton mode collectively. The coupled system can be described in a six-component basis $\Phi=(|P^+>,|\uparrow^+>,|\downarrow^+>,|P^->,|\uparrow^->,|\downarrow^->)$, 
%%$\Phi=(|P^+>,|K^+>,|K'^+>,|P^->,|K^->,|K'^->)$
%where $|P^{\pm}>$ are the two degenerate eigenmodes of the rotation operator of the PhC, $|\uparrow^{\pm}>$ and  $|\downarrow^{\pm}>$ are the collective excitonic modes with opposite spins and couple to $|P^{\pm}>$, respectively. 
The effective Hamiltonian is:
\begin{equation}
    H_{\mathbf{k}}=
    \begin{bmatrix}
        %\omega_{p1} & \alpha &\beta & \nu(k_x-ik_y)^2 &0 &0\\ \alpha^* & \omega_{K} &0 &0 &0 &0\\ \beta^* & 0 &\omega_{K'} &0 &0 &0\\ \nu(k_x+ik_y)^2 &0 &0 &\omega_{p2} &\beta^* &\alpha^*\\      0 &0 &0 &\beta &\omega_{K} &0\\       0 &0 &0 &\alpha &0 &\omega_{K'}\\
        %\mu_1 k^2 & \alpha &\beta & \nu k^2e^{-i2\phi_{\mathbf{k}}} &0 &0\\ \alpha^* & \omega_{\uparrow} &0 &0 &0 &0\\ \beta^* & 0 &\omega_{\downarrow} &0 &0 &0\\ \nu k^2e^{i2\phi_{\mathbf{k}}} &0 &0 &-\mu_2 k^2 &\beta^* &\alpha^*\\      0 &0 &0 &\beta &\omega_{\uparrow} &0\\       0 &0 &0 &\alpha &0 &\omega_{\downarrow}\\
        \omega_{\mathbf{k}}^{P} & \alpha &\beta & \nu k^2e^{-i2\phi_{\mathbf{k}}} &0 &0\\ \alpha^* & \omega_{\uparrow} &0 &0 &0 &0\\ \beta^* & 0 &\omega_{\downarrow} &0 &0 &0\\ \nu k^2e^{i2\phi_{\mathbf{k}}} &0 &0 &\omega_{\mathbf{k}}^{P} &\beta^* &\alpha^*\\      0 &0 &0 &\beta &\omega_{\uparrow} &0\\       0 &0 &0 &\alpha &0 &\omega_{\downarrow}\\
    \end{bmatrix}.
    \label{eq:Hk_quadratic}
\end{equation}
Here, $\omega_{\mathbf{k}}^{P}$ is the eigenvalue of $|P^\pm>$% and $|P^->$
and $\omega_{k=\Gamma}^{P}=0$, $\omega_{\uparrow}$ and $\omega_{\downarrow}$ are the energies of exciton modes coupled to $|P^+>$ and $|P^->$, respectively. 
%\xin{Do we need to clarify that $\omega_{\mathbf{k}}^{\pm}$ are identical for any $\mathbf{k}$? }
%\xin{In a simplified model, $\omega_{\mathbf{k}}^p$ for mode $|P^+>$ and $|P^->$ are supposed to be 0. The quadratic characteristic results from the coupling term. This is the case introduced in Fig.~\ref{f3}(a). For specific examples, $\omega_{\mathbf{k}}^p$ could be diverse functions of k and supposed to be identical constrained by the symmetry.}
$\alpha$ and $\beta$ are the corresponding collective coupling strengths, influenced by the spatial distribution of the PhC mode profiles. 
%\hui{Is this correct now?}
% \xin{the current one represent the case in Fig.~\ref{f_sch_quad}(a). If we don't show the result in current Fig. \ref{f4sim_qua_TM}(b-c), the current one works.}
% \hui{Let's discuss after checking other comments?}
% \xin{Sure}

%Breaking TR symmetry via the excitonic component results in a gapped polariton spectrum with a non-trivial bulk topology. 

Consider the simplest example of $\omega_{\mathbf{k}}^{P}=0$, the band structure of the above Hamiltonian is shown in Figs.~\ref{f_sch_quad}(a-c) for three different scenarios: with degenerate excitons modes that preserve TR symmetry, and with a single or two non-degenerate exciton modes that both break TR symmetry. 
%Figures \ref{f_sch_quad} (b-d) illustrate polariton dispersion in various coupling scenarios: with degenerate excitons, a single exciton, and split excitons. 
Without excitons, two quadratic photonic bands with touching at $\Gamma$ are formed due to winding coupling between the degenerate PhC modes, as indicated by the black dashed curves in Figs. \ref{f_sch_quad}(a-c)). With excitons but under TR symmetry, $\omega_{\uparrow}=\omega_{\downarrow}$, strong exciton-photon coupling leads to two pairs of quadratic-touching, bright polariton bands and one pair of degenerate exciton bands without photonic components, as shown in Fig. \ref{f_sch_quad}(a). Degeneracies at $\Gamma$ remain due to symmetry protection. %The four bright polariton bands are doubly degenerate at the $\Gamma$ point, forming upper and lower polaritonic touching bands. 
TR symmetry is broken when $\omega_{\uparrow}\neq\omega_{\downarrow}$, which lifts all degeneracy at $\Gamma$ and results in gapped topological polariton bands with nontrivial Chern numbers. 
Figure \ref{f_sch_quad}(b) shows the result when one of the exciton modes is very far detuned or completely saturated. The Hamiltonian reduces to two photon modes coupled with a single bright exciton mode, resulting in four topological polariton bands with two topological bandgaps, highlighted as red and blue stripes in Fig. \ref{f_sch_quad}(b). The Chern numbers for each band are confirmed as $\pm1$, with the signs indicated by blue and red. 
Figure \ref{f_sch_quad}(c) shows the result when spin-up and spin-down excitons are split by $\Delta\omega=\omega_{\uparrow}-\omega_{\downarrow}$, yielding six topological polariton bands with three topological bandgaps. This configuration is more common in experimental systems, where $\Delta\omega$ can be introduced by Zeeman splitting or the optical Stark effect. Note that dark excitons in this system are inside the trivial gaps and would not affect the topological gaps or edge states. This is different from the first type based on BIC, where the topological gap needs to be opened at the crossing of the exciton and photon bands.

%\textcolor{red}{
Opening the gap at $\Gamma$ is also accompanied by a different distribution of the Berry curvature compared to those of Dirac cone bands. The Berry curvature of the gapped quadratic touching bands forms a ring around $\Gamma$ (Fig.~\ref{f_sch_quad}(d)), in contrast to a sharp peak around the Dirac cones (Fig.~\ref{f_sch_quad}(e)). The relatively isotropic and broad distribution of the Berry curvature may facilitate studies of the quantum Hall effect in such systems and is known to help stabilize fractional excitations.

\section{The topological gap}
%Despite the differences between Dirac cones and quadratic touching bands in terms of winding coupling terms and momentum space distribution, they share similarities in their gapped spectra, particularly in the topological gap. 
%The topological gap size in such polariton system, exhibit minimum at the original degenerate point ($K/K'$ for Dirac cones, $\Gamma$ for quadratic touching bands). At these points, the winding coupling terms between two photonic bands vanish, leading to the same Hamiltonian for both cases. Consequently, they share common parameters for optimizing the topological gap size, including exciton-photon detuning, coupling strengths, and Zeeman splitting.

\begin{figure}
\centering
\includegraphics[width=1\linewidth]{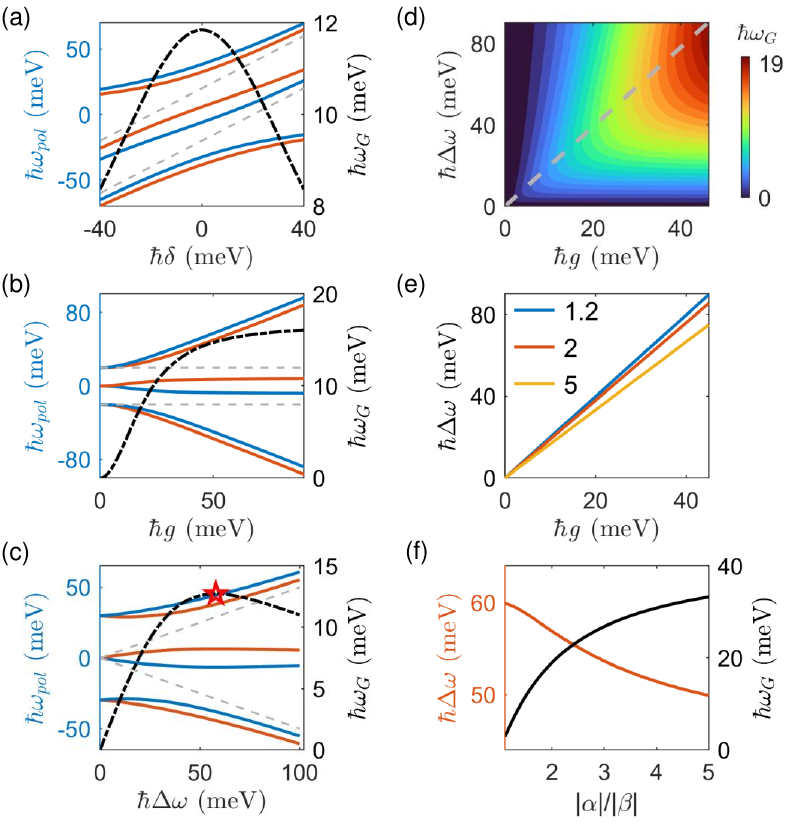}
\caption{\label{f3} (a-c) Eigenvalues $\omega_{pol}$ of the six polariton bands (blue and red solid lines) measured from the original degenerate point, and the middle gap $\omega_G$ (black dashed lines) between the third and forth modes as a function of (a) detuning $\delta$ with a fixed Zeeman splitting of 40~meV and $\hbar g=30$~meV; (b) total coupling strength $g$ with a fixed Zeeman splitting of 40~meV; and (c) Zeeman splitting $\Delta\omega$ with a fixed $\hbar g = 30 $~meV. The gray dashed lines represent the energy of two split excitons. The red star in (c) denotes the optimal Zeeman splitting to achieve the maximum topological gap. (d) Contour plot of the topological gap as a function of the total coupling strength $g$ and Zeeman splitting $\Delta\omega$. The dashed line represents the optimal Zeeman splitting with maximum gap for each $g$. (e) Optimal Zeeman splitting as a function of $g$ for varying values of $|\alpha|/|\beta|$. (f) Optimal Zeeman splitting $\Delta\omega$ (red line) and maximum gap (black line) as a function of $|\alpha|/|\beta|$ with a constant $\hbar g = 30$~meV. 
%\hui{I meant fist column a-c, second column d-f. :-) Also, need to add $\hbar$ where meV is used as the unit.}\xin{done}
} 
\end{figure}
\begin{figure}
\centering
\includegraphics[width=1\linewidth]{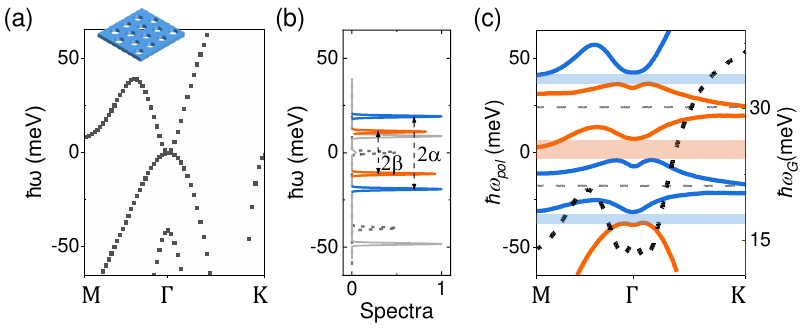}
\caption{\label{f4sim_qua_TM} 
%(a) Band structure of a 2D GaP PhC with quadratic touching at $\Gamma$. Inset shows a schematic of the PhC made of a triangular lattice of circular air holes. (b) Polariton bands under preserved TR symmetry, where the quadratic touching point is resonant with two degenerate exciton states. (c,d) Polariton Chern bands in the TR symmetry-breaking case, obtained by (c) solving the eigenvalues of the effective Hamiltonian and (d) numerical simulation using FDTD. Horizontal dashed lines indicate the two valley exciton states with an optimal valley exciton splitting of 42.5~meV. Dashed black lines in (c) represent the uncoupled case. Blue/red solid lines represent polariton bands with Chern numbers $\pm 1$. The black dashed line shows the middle gap size as a function of $k$.
Numerical simulation of band structures of a 2D GaP PhC-TMD systems. (a) Pure PhC bands with quadratic touching at $\Gamma$. Inset shows a schematic of the PhC made of a triangular lattice of circular air holes. (b) Simulated spectra from pure PhC mode (dashed lines) and cavity with single valley exciton (solid lines) at $\Gamma$ point. $\alpha$ and $\beta$ can be determined by the split four polariton peaks illustrated by the blue and red lines. (c) Polariton Chern bands (blue and red solid lines for Chern number $\pm 1$) when the PhC bands in (a) are coupled with two valley exciton states with $\delta=0$ and $\hbar\Delta\omega=42.5$~meV splitting (black dotted lines). The blue and red strips mark the gaps between two Chern bands of different Chern numbers. The black dashed line shows the middle gap size as a function of $\mathbf{k}$.
%\hui{I'm thinking we may not need (b-c), only keep the simulation results (a) and (d). Possibly combine them with Fig. 3.}
%\xin{done}
%\hui{sub figure (c) is perhaps too small. Perhaps we just keep (a) and (c) after all.}\xin{I revised the size for each subfigure.}
%\hui{Perhaps use the same y-axis range for all sub figures, say -75 to 75 meV. Also, for all figures, the font size -- (a) labels, tick labels etc -- preferably should be same or a little smaller than the font size of the caption. But can wait till we finalized all contents to modify them.}\xin{revised accordingly.}
} 
\end{figure}
\begin{figure*}
\centering
\includegraphics[width=1\linewidth]{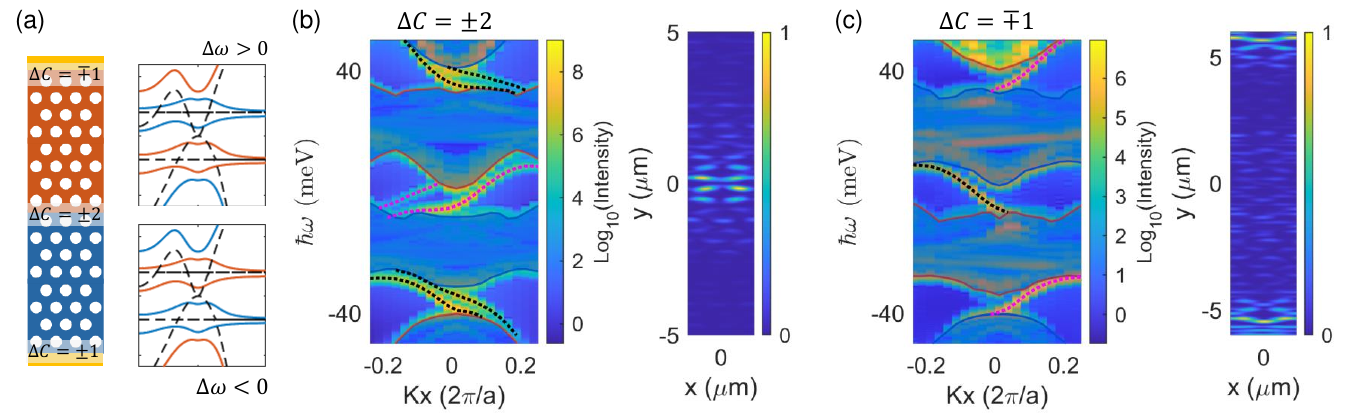}
\caption{\label{sf2} (a) Schematic of 1D interfaces with $|\Delta C| = 2$ and 1, formed by two regions with opposite valley exciton splitting (orange and blue areas), and regions between topological polariton and trivial metal (yellow area). The right panel shows the corresponding bandstructure of the topological polaritons band, with the color representing the sign of Chern number. (b) Dispersion of the chiral edge states (dashed lines) within three topological gaps with $\Delta C = \pm2$, and an example of mode profile of edge state. (c) Dispersion of the chiral edge states (dashed lines) within three topological gaps with $\Delta C = \pm1$, and an example of mode profile of the edge state. }
\end{figure*}
The size of the topological gap is a key metric of topological protection. Here, we use the quadratic touching type bands to examine how the topological gaps %induced by split excitons (Fig.~\ref{f_sch_quad}(d)) 
depend on the exciton-photon detuning $\delta=(\omega_\uparrow+\omega_\downarrow)/2$, 
total exciton-photon coupling $g=\sqrt{|\alpha|^2+|\beta|^2}$, 
ratio of exciton-photon coupling with the two bands $|\alpha|/|\beta|$, and the strength of TR breaking $\Delta\omega$. The results are shown in Fig. \ref{f3}. 
%\hui{Use the same symbol as in (1). For now, replaced $\Delta_Z$ by $\Delta\omega$}. 

As shown in Fig.~\ref{f3}(a), the gaps show maxima at zero detuning $\delta=0$. The largest is the middle gap, between the 3rd and 4th band, reaching close to 12~meV for the parameters used ($\alpha / \beta = 1.56$, $\hbar g=30$~meV and $\hbar\Delta\omega =40$~meV). The results reflect that the gap is optimized with maximal exciton-photon hybridization.
%The blue and red solid lines show the evolution of the six eigenvalues at $\Gamma$, while the black dashed lines represent changes in the central gap.
%The topological gap exhibits distinct dependencies on these parameters. Figure~\ref{f3}(a) shows that as the absolute detuning $|\delta|$ decreases, the gap increases and reaches its maximum at $\delta=0$. 
Fixing at $\delta=0$, the gaps increase with increasing total coupling strength $g$, but start to saturate at $g\sim\hbar\Delta\omega=40$~meV (Fig.~\ref{f3}(b)). The saturation can be understood as due to the finite TR symmetry breaking $\Delta\omega$. Similarly, with fixed $\delta=0$ and $\hbar g=30$~meV, the gaps increase as $\Delta\omega$ increases from zero, reach maxima at $\Delta\omega\sim 1.93g$ and gradually decrease at larger $\Delta\omega$, as the exciton-photon hybridization decreases and the modes become either exciton- or photon-like.

The above results reflect that the topological properties of the system arise from both TR symmetry breaking and exciton-photon hybridization. The gap size critically depends on not only the strength of TR symmetry breaking ($\Delta\omega$), but also the strength of the exciton-photon coupling ($g$) and the amount of exciton-photon hybridization.  %This saturation occurs because the fixed exciton splitting limits the strength of TR symmetry breaking. Conversely, as $\Delta_Z$ increases, the gap reaches a peak value of around 12~meV at $\hbar\Delta_Z = 58$~meV (indicated by the red star in Fig.~\ref{f3}), before declining with further increases in Zeeman splitting.
%
%The evolution of the gap with $\Delta_Z$ results from a trade-off between increased TR symmetry breaking from Zeeman splitting and reduced symmetry breaking due to decreased coupling strength as exciton-photon detuning increases. These findings highlight the importance of balancing coupling strength $g$ and Zeeman splitting for optimal topological gap performance.
%\hui{I'm not sure how to understand the above commented out statement. If you are refering to the exciton fraction in the modes, the upper and lower branches have increasing exciton fractions with increasing g, but also show saturation in the sizes of the corresponding upper and lower gaps... Could you clarify or check?}
%\xin{As Zeeman splitting increase, the detuning between exciton and photon increase, leading to a reduction in the excitonic fraction. Even though Zeeman splitting increase, the resulted splitting of polariton reduce. More concisely, the evolution of the gap with $\Delta\omega$ results from a trade-off between increased TR symmetry breaking from Zeeman splitting and reduced excitonic fraction as exciton-photon detuning increases.}
%\xin{In this sense, $\Delta\omega$ in (1) refer to the gap. In TMD polariton system, the gap is determined by $\alpha,\beta$ and Zeeman splitting. Maybe we could use $\Delta_z$ represent zeeman splitting, but $\Delta\omega_{pol}$ for topological gap. Please let me know what you think. }
%
In Fig.~\ref{f3}(d), we analyze how the optimal middle gap size depends on the relation between $g$ and $\Delta\omega$, with $\delta = 0$. The gray dashed line traces optimal $g$ at each $\Delta\omega$ for the maximum gap size, revealing a nearly linear dependence within the range shown. Interestingly, the slope of the linear relationship depends on the difference between $|\alpha|$ and $|\beta|$. As shown in Fig.~\ref{f3}(e), the slope decreases with increasing imbalance between $|\alpha|$ and $|\beta|$. This suggests that, given $g$, less TR symmetry breaking is needed for a larger gap when the difference between $|\alpha|$ and $|\beta|$ increases, as we show in Fig.~\ref{f3}(f) (orange line) for $\hbar g=30$~meV. Fig.~\ref{f3}(f) also shows that, for given $g$ and optimal $\Delta\omega$, the middle gap increases with increasing $|\alpha|/|\beta|$, saturating to a value slightly above $g$ (black line).

These results highlight key factors for optimizing the topological gap, including a large difference in the coupling strengths $|\alpha|$ and $|\beta|$, a large total coupling strength $g$, strong exciton-photon mixing measured by exciton-photon detuning, and strong TR symmetry breaking measured by $\Delta\omega$. 
\section{Practical realization of a large topological gap}
The potential to obtain a large topological gap across the BZ provides an exciting prospect for experimental realization and applications of topological polariton systems. 
%improving the visibility and stability of topological features for experimental observation.
Below we provide an example design based on practical experimental systems. 

To achieve a large topological gap, the coupled system of 2D PhC and monolayer transition metal dichalcogenides (TMDs) serves as a promising candidate. 
2D PhCs allow the design freedom to achieve a large $|\alpha|/|\beta|$ ratio, and to enhance the total coupling strength $g$ by increasing photon confinement and exciton-photon overlap. Semiconductor TMD monolayers can be directly and efficiently integrated with PhCs~\cite{Zhang_Photoniccrystal_2018a}. Moreover, they feature both large oscillator strengths that enables both a large $g$ \cite{Liu_Generation_2020b, Li_Experimental_2021a} and a large $\Delta\omega$ through the valley-selective optical Stark effect~\cite{Cunningham_Resonant_2019,Yong_Valleydependent_2019,Zhou_Cavity_2024}. 
%increasing exciton density~\cite{Zhao_Exciton_2023} or improving photonic confinement, and the optimal Zeeman splitting can be determined accordingly. 
%Strong exciton-photon interaction in TMDs facilitates larger coupling strengths~\cite{Zhang_Photoniccrystal_2018a, Liu_Generation_2020b, Li_Experimental_2021a}. While magnetic fields can induce only limited splitting (a few meV) even at strengths of tens of Tesla~\cite{Li_Valley_2014, Aivazian_Magnetic_2015, MacNeill_Breaking_2015, Srivastava_Valley_2015}, the valley-selective optical Stark effect in TMDs can achieve significant valley exciton splittings up to tens of meV~\cite{Cunningham_Resonant_2019, Yong_Valleydependent_2019}, providing a more effective approach to breaking TR symmetry. 
Additionally, the rich excitonic physics in 2D materials, such as moiré excitons, provides further opportunities to break TR symmetry~\cite{Mak_Semiconductor_2022a, Chang_Colloquium_2023, Zhao_Realization_2024a}. 
We consider the example of a GaP PhC consisting of a triangular lattice of circular air holes, as shown in Fig.~\ref{f4sim_qua_TM}(a). The band structures are obtained by Lumerical FDTD and feature quadratic touching bands at TMD exciton frequencies. To form polaritons, a TMD monolayer is placed on top of the PhC as a dielectric material with exciton resonances as Lorentz holes~\cite{Li_Measurement_2014} in the permittivity tensor. 
%From the simulated spectrum from the cavity coupled to a single valley exciton shown in Fig. \ref{f4sim_qua_TM}(b), we extract $\hbar g = 22~meV$, $|\alpha|/|\beta|=1.73$.
When there is only one valley exciton, with $\delta=0$, we obtain two pairs of upper and lower polaritons, as shown in Fig.~\ref{f4sim_qua_TM}(b), similar to Fig.~\ref{f_sch_quad}(c). From the polariton splittings, we extract $\hbar g = 22~meV$, $|\alpha|/|\beta|=1.73$.
%strong coupling leads to six bands in three pairs -- the upper polaritons, excitons, and lower polaritons, all topologically trivial and each pair preserving the symmetry-protected degeneracy at $\Gamma$. 
%\hui{It will be good to add a simulation figure that we use to obtain g and the coupling ratio. Please update the above description as needed.}
%\xin{we don't have simulated results about TR symmetry preserved case. we can model it using the effective Hamiltonian. Shall we add this part of data?}
%\hui{That's OK. How about using similar simulated spectral data as in Fig. 3c, if you have it. Otherwise current Fig. 5b is also OK, but perhaps color code the peaks? Also it may be better to use meV for y-axis in (a) to make it easier to compare with b-c.}
%\xin{done}
When there are two exciton bands split by $\hbar\Delta\omega=$42.5~meV (Fig.~\ref{f4sim_qua_TM}(c)), six polariton Chern bands with Chern number of $\pm1$ are formed with three topological gaps (blue and red areas) and two trivial gaps. The middle gap is about 12~meV wide. %The Chern number of each band is confirmed by \hui{...}
%\xin{the Chern number here is confirmed by the effective Hamiltonian and further confirmed by the number of chiral edge states. Since we mainly refer to numerical simulation results here, shall we remove this statement?}
These results fully agree with the analysis of the effective Hamiltonian in Eq.~\ref{eq:Hk_quadratic} and Fig.~\ref{f3}. 
%Figures \ref{f4sim_qua_TM} (b-c) display the polariton band structure for cases with preserved and broken TR symmetry, respectively. These results are obtained by solving the eigenvalues of the effective Hamiltonian with constant coupling strength, derived from simulations at $\Gamma$ point, where  
%The introduction of an optimal valley splitting of 42.5~meV results in a gapped spectrum with a gap size of approximately 12~meV, as shown in Fig. \ref{f4sim_qua_TM}(c). It agrees well with the FDTD simulation results in Fig. \ref{f4sim_qua_TM}(d). 

%\subsubsection{Edge state}
The large topological gaps obtained in our example already exceed the reported TMD polariton linewidths. The system should support robust chiral edge modes around $\Gamma$ that are readily observable in experiments. We verify such edge modes in our simulation. As shown in Fig.~\ref{sf2}(a), we construct 1D edges with $|\Delta C| = 2$ and $1$. The interface with $|\Delta C| = 2$ is formed by two TMD-PhC regions with opposite exciton splitting $\Delta \omega$, resulting in identical band structures that have opposite Chern numbers. %Consequently, this interface corresponds to a change in Chern number $\Delta C = \pm2$ across the three gaps, leading to the emergence of 
Two chiral edge bands are observed in each of the three topological gaps, as shown in Fig.~\ref{sf2}(b). For comparison, we also form an interface between topological polariton regions and trivial metal with $|\Delta C| = 1$. Correspondingly, one chiral edge mode is formed in each band gap, as shown in Fig.~\ref{sf2}(c). Notably, in both cases, the middle gap displays an opposite $\Delta C$ compared to the upper and lower gaps; correspondingly, the chiral edge states have opposite chirality as expected. The mode profiles shown in the right panels of Figs.~\ref{sf2}(b) and (c) confirm strong localization of the edge states. These results align well with the bulk-edge correspondence in topological insulators. The topologically protected chiral edge modes have significant potential for nanophotonic devices applications. 

%Numerical simulations further confirm the presence of chiral edge states around $\Gamma$ (see Supplementary Material). The large gap of $12$~meV across the BZ surpasses the polariton linewidth and supports robust chiral edge modes around $\Gamma$, as we also confirm with numerical simulations (see Supplementary Material), and should be readily observable in experiment. 
%improving the visibility and stability of topological features for experimental observation.

%The widespread occurrence of quadratic touching bands, combined with the high design flexibility of 2D PhCs, offers extensive opportunities for designing topological systems beyond Dirac cones, enabling the achievement of substantial complete bandgap. Additionally, leveraging quadratic touching bands around $\Gamma$ simplifies experimental access to topological features without further lattice engineering.
%

\section{Conclusion}
In summary, we show that 2D PhC-polaritons provide a versatile platform to realize Chern bands with large topological gaps at the optical frequencies. Beyond the conventional Dirac point framework, we show two types of common 2D PhC bands as effective platforms for achieving polariton Chern bands: those with symmetry-protected BICs and with symmetry-protected quadratic touching points. 
Both types of bands commonly exist in 2D PhCs of C3, C4 and C6 symmetries, and the topological gap opens at the $\Gamma$ point, providing improved design flexibility and experimental accessibility. The ring-shaped Berry curvature distribution may facilitate experimental studies of the quantum Hall effect. Bands with higher Chern numbers are readily obtained around BIC in C6 PhCs. Furthermore, we show key factors that determine the size of the topological gap at symmetry-protected quadratic touch degeneracy, and we provide a practical design with a topological gaps over 10~meV, two orders of magnitude greater than previous experientially feasible schemes. 
%We also show that the topological gap in quadratic touching bands can be maximized by increasing the difference of the exciton-photon coupling with the two photon bands, aligning the mean exciton resonance to the original photonic degeneracy point, and increasing the exciton-photon coupling and TR symmetry breaking strengths together proportionally. 
These findings underscore the broad applicability of polariton Chern bands in 2D PhCs, expand the design space for topological photonic systems, and provide guides for achieving improved or novel topological properties~\cite{Skirlo_Multimode_2014,Lu_Topological_2016,Wu_Applications_2017}. 
% \hui{Could be helpful to add here some references, like theoretical proposals, on applications of photonic Chern bands and higher order Chern bands.}
% \xin{\cite{Skirlo_Multimode_2014} shows the application of higher Chern number on multimode one-way waveguide using gyroelectric photonic crystals. \cite{Lu_Topological_2016} provides a general review including a highlight on their previous work about higher Chern bands~\cite{Skirlo_Multimode_2014}. \cite{Wu_Applications_2017} is a review about topological photonics in integrated photonic devices}

%We revealed that the optimization of the topological gap is critically dependent on photonic mode profiles and exciton splitting. Practical designs based on quadratic touching bands predicts a larger gap about 12~meV, enabling the robust chiral edge state. Moreover, our study highlights the capability of BICs on the realization of higher Chern number bands, facilitating the discovery of exotic state and phenomena. These result underscores the broad applicability of polariton Chern bands in 2D PhCs, providing new avenues for the development of advanced photonic devices with superior performance and diverse functionalities. Our work paves the way for future experimental and theoretical studies to further explore and harness the potential of topological photonics, ultimately contributing to the advancement of next-generation photonic technologies.
\section*{Acknowledgment}
X.X. and H.D. acknowledge the support by the Army Research Office under Awards W911NF-17-1-0312, the Air Force Office of Scientific Research under Awards FA2386-21-1-4066, the National Science Foundation under Awards DMR 2132470, the Office of Naval Research under Awards N00014-21-1-2770, and the Gordon and Betty Moore Foundation under Grant GBMF10694. X.X. acknowledges the support by Schmidt Sciences, LLC.

\bibliography{maintext}% Produces the bibliography via BibTeX.

\begin{thebibliography}{75}%
\makeatletter
\providecommand \@ifxundefined [1]{%
 \@ifx{#1\undefined}
}%
\providecommand \@ifnum [1]{%
 \ifnum #1\expandafter \@firstoftwo
 \else \expandafter \@secondoftwo
 \fi
}%
\providecommand \@ifx [1]{%
 \ifx #1\expandafter \@firstoftwo
 \else \expandafter \@secondoftwo
 \fi
}%
\providecommand \natexlab [1]{#1}%
\providecommand \enquote  [1]{``#1''}%
\providecommand \bibnamefont  [1]{#1}%
\providecommand \bibfnamefont [1]{#1}%
\providecommand \citenamefont [1]{#1}%
\providecommand \href@noop [0]{\@secondoftwo}%
\providecommand \href [0]{\begingroup \@sanitize@url \@href}%
\providecommand \@href[1]{\@@startlink{#1}\@@href}%
\providecommand \@@href[1]{\endgroup#1\@@endlink}%
\providecommand \@sanitize@url [0]{\catcode `\\12\catcode `\$12\catcode
  `\&12\catcode `\#12\catcode `\^12\catcode `\_12\catcode `\%12\relax}%
\providecommand \@@startlink[1]{}%
\providecommand \@@endlink[0]{}%
\providecommand \url  [0]{\begingroup\@sanitize@url \@url }%
\providecommand \@url [1]{\endgroup\@href {#1}{\urlprefix }}%
\providecommand \urlprefix  [0]{URL }%
\providecommand \Eprint [0]{\href }%
\providecommand \doibase [0]{https://doi.org/}%
\providecommand \selectlanguage [0]{\@gobble}%
\providecommand \bibinfo  [0]{\@secondoftwo}%
\providecommand \bibfield  [0]{\@secondoftwo}%
\providecommand \translation [1]{[#1]}%
\providecommand \BibitemOpen [0]{}%
\providecommand \bibitemStop [0]{}%
\providecommand \bibitemNoStop [0]{.\EOS\space}%
\providecommand \EOS [0]{\spacefactor3000\relax}%
\providecommand \BibitemShut  [1]{\csname bibitem#1\endcsname}%
\let\auto@bib@innerbib\@empty
%</preamble>
\bibitem [{\citenamefont {Lu}\ \emph {et~al.}(2014)\citenamefont {Lu},
  \citenamefont {Joannopoulos},\ and\ \citenamefont {Solja{\v
  c}i{\'c}}}]{Lu_Topological_2014}%
  \BibitemOpen
  \bibfield  {author} {\bibinfo {author} {\bibfnamefont {L.}~\bibnamefont
  {Lu}}, \bibinfo {author} {\bibfnamefont {J.~D.}\ \bibnamefont
  {Joannopoulos}},\ and\ \bibinfo {author} {\bibfnamefont {M.}~\bibnamefont
  {Solja{\v c}i{\'c}}},\ }\bibfield  {title} {\bibinfo {title} {Topological
  photonics},\ }\href {https://doi.org/10.1038/nphoton.2014.248} {\bibfield
  {journal} {\bibinfo  {journal} {Nat. Photonics}\ }\textbf {\bibinfo {volume}
  {8}},\ \bibinfo {pages} {821} (\bibinfo {year} {2014})}\BibitemShut {NoStop}%
\bibitem [{\citenamefont {Khanikaev}\ and\ \citenamefont
  {Shvets}(2017)}]{Khanikaev_Twodimensional_2017}%
  \BibitemOpen
  \bibfield  {author} {\bibinfo {author} {\bibfnamefont {A.~B.}\ \bibnamefont
  {Khanikaev}}\ and\ \bibinfo {author} {\bibfnamefont {G.}~\bibnamefont
  {Shvets}},\ }\bibfield  {title} {\bibinfo {title} {Two-dimensional
  topological photonics},\ }\href {https://doi.org/10.1038/s41566-017-0048-5}
  {\bibfield  {journal} {\bibinfo  {journal} {Nat. Photonics}\ }\textbf
  {\bibinfo {volume} {11}},\ \bibinfo {pages} {763} (\bibinfo {year}
  {2017})}\BibitemShut {NoStop}%
\bibitem [{\citenamefont {Ozawa}\ \emph {et~al.}(2019)\citenamefont {Ozawa},
  \citenamefont {Price}, \citenamefont {Amo}, \citenamefont {Goldman},
  \citenamefont {Hafezi}, \citenamefont {Lu}, \citenamefont {Rechtsman},
  \citenamefont {Schuster}, \citenamefont {Simon}, \citenamefont {Zilberberg},\
  and\ \citenamefont {Carusotto}}]{Ozawa_Topological_2019}%
  \BibitemOpen
  \bibfield  {author} {\bibinfo {author} {\bibfnamefont {T.}~\bibnamefont
  {Ozawa}}, \bibinfo {author} {\bibfnamefont {H.~M.}\ \bibnamefont {Price}},
  \bibinfo {author} {\bibfnamefont {A.}~\bibnamefont {Amo}}, \bibinfo {author}
  {\bibfnamefont {N.}~\bibnamefont {Goldman}}, \bibinfo {author} {\bibfnamefont
  {M.}~\bibnamefont {Hafezi}}, \bibinfo {author} {\bibfnamefont
  {L.}~\bibnamefont {Lu}}, \bibinfo {author} {\bibfnamefont {M.~C.}\
  \bibnamefont {Rechtsman}}, \bibinfo {author} {\bibfnamefont {D.}~\bibnamefont
  {Schuster}}, \bibinfo {author} {\bibfnamefont {J.}~\bibnamefont {Simon}},
  \bibinfo {author} {\bibfnamefont {O.}~\bibnamefont {Zilberberg}},\ and\
  \bibinfo {author} {\bibfnamefont {I.}~\bibnamefont {Carusotto}},\ }\bibfield
  {title} {\bibinfo {title} {Topological photonics},\ }\href
  {https://doi.org/10.1103/RevModPhys.91.015006} {\bibfield  {journal}
  {\bibinfo  {journal} {Rev. Mod. Phys.}\ }\textbf {\bibinfo {volume} {91}},\
  \bibinfo {pages} {015006} (\bibinfo {year} {2019})}\BibitemShut {NoStop}%
\bibitem [{\citenamefont {Smirnova}\ \emph {et~al.}(2020)\citenamefont
  {Smirnova}, \citenamefont {Leykam}, \citenamefont {Chong},\ and\
  \citenamefont {Kivshar}}]{Smirnova_Nonlinear_2020}%
  \BibitemOpen
  \bibfield  {author} {\bibinfo {author} {\bibfnamefont {D.}~\bibnamefont
  {Smirnova}}, \bibinfo {author} {\bibfnamefont {D.}~\bibnamefont {Leykam}},
  \bibinfo {author} {\bibfnamefont {Y.}~\bibnamefont {Chong}},\ and\ \bibinfo
  {author} {\bibfnamefont {Y.}~\bibnamefont {Kivshar}},\ }\bibfield  {title}
  {\bibinfo {title} {Nonlinear topological photonics},\ }\href
  {https://doi.org/10.1063/1.5142397} {\bibfield  {journal} {\bibinfo
  {journal} {Appl. Phys. Rev.}\ }\textbf {\bibinfo {volume} {7}},\ \bibinfo
  {pages} {021306} (\bibinfo {year} {2020})}\BibitemShut {NoStop}%
\bibitem [{\citenamefont {Kim}\ \emph {et~al.}(2020)\citenamefont {Kim},
  \citenamefont {Jacob},\ and\ \citenamefont {Rho}}]{Kim_Recent_2020}%
  \BibitemOpen
  \bibfield  {author} {\bibinfo {author} {\bibfnamefont {M.}~\bibnamefont
  {Kim}}, \bibinfo {author} {\bibfnamefont {Z.}~\bibnamefont {Jacob}},\ and\
  \bibinfo {author} {\bibfnamefont {J.}~\bibnamefont {Rho}},\ }\bibfield
  {title} {\bibinfo {title} {Recent advances in {{2D}}, {{3D}} and higher-order
  topological photonics},\ }\href {https://doi.org/10.1038/s41377-020-0331-y}
  {\bibfield  {journal} {\bibinfo  {journal} {Light Sci. Appl.}\ }\textbf
  {\bibinfo {volume} {9}},\ \bibinfo {pages} {130} (\bibinfo {year}
  {2020})}\BibitemShut {NoStop}%
\bibitem [{\citenamefont {Umucal{\i}lar}\ and\ \citenamefont
  {Carusotto}(2011)}]{Umucalilar_Artificial_2011}%
  \BibitemOpen
  \bibfield  {author} {\bibinfo {author} {\bibfnamefont {R.~O.}\ \bibnamefont
  {Umucal{\i}lar}}\ and\ \bibinfo {author} {\bibfnamefont {I.}~\bibnamefont
  {Carusotto}},\ }\bibfield  {title} {\bibinfo {title} {Artificial gauge field
  for photons in coupled cavity arrays},\ }\href
  {https://doi.org/10.1103/PhysRevA.84.043804} {\bibfield  {journal} {\bibinfo
  {journal} {Phys. Rev. A}\ }\textbf {\bibinfo {volume} {84}},\ \bibinfo
  {pages} {043804} (\bibinfo {year} {2011})}\BibitemShut {NoStop}%
\bibitem [{\citenamefont {Hafezi}\ \emph {et~al.}(2011)\citenamefont {Hafezi},
  \citenamefont {Demler}, \citenamefont {Lukin},\ and\ \citenamefont
  {Taylor}}]{Hafezi_Robust_2011}%
  \BibitemOpen
  \bibfield  {author} {\bibinfo {author} {\bibfnamefont {M.}~\bibnamefont
  {Hafezi}}, \bibinfo {author} {\bibfnamefont {E.~A.}\ \bibnamefont {Demler}},
  \bibinfo {author} {\bibfnamefont {M.~D.}\ \bibnamefont {Lukin}},\ and\
  \bibinfo {author} {\bibfnamefont {J.~M.}\ \bibnamefont {Taylor}},\ }\bibfield
   {title} {\bibinfo {title} {Robust optical delay lines with topological
  protection},\ }\href {https://doi.org/10.1038/nphys2063} {\bibfield
  {journal} {\bibinfo  {journal} {Nat. Phys.}\ }\textbf {\bibinfo {volume}
  {7}},\ \bibinfo {pages} {907} (\bibinfo {year} {2011})}\BibitemShut {NoStop}%
\bibitem [{\citenamefont {Hafezi}\ \emph {et~al.}(2013)\citenamefont {Hafezi},
  \citenamefont {Mittal}, \citenamefont {Fan}, \citenamefont {Migdall},\ and\
  \citenamefont {Taylor}}]{Hafezi_Imaging_2013}%
  \BibitemOpen
  \bibfield  {author} {\bibinfo {author} {\bibfnamefont {M.}~\bibnamefont
  {Hafezi}}, \bibinfo {author} {\bibfnamefont {S.}~\bibnamefont {Mittal}},
  \bibinfo {author} {\bibfnamefont {J.}~\bibnamefont {Fan}}, \bibinfo {author}
  {\bibfnamefont {A.}~\bibnamefont {Migdall}},\ and\ \bibinfo {author}
  {\bibfnamefont {J.~M.}\ \bibnamefont {Taylor}},\ }\bibfield  {title}
  {\bibinfo {title} {Imaging topological edge states in silicon photonics},\
  }\href {https://doi.org/10.1038/nphoton.2013.274} {\bibfield  {journal}
  {\bibinfo  {journal} {Nat. Photonics}\ }\textbf {\bibinfo {volume} {7}},\
  \bibinfo {pages} {1001} (\bibinfo {year} {2013})}\BibitemShut {NoStop}%
\bibitem [{\citenamefont {Wu}\ and\ \citenamefont {Hu}(2015)}]{Wu_Scheme_2015}%
  \BibitemOpen
  \bibfield  {author} {\bibinfo {author} {\bibfnamefont {L.-H.}\ \bibnamefont
  {Wu}}\ and\ \bibinfo {author} {\bibfnamefont {X.}~\bibnamefont {Hu}},\
  }\bibfield  {title} {\bibinfo {title} {Scheme for {{Achieving}} a
  {{Topological Photonic Crystal}} by {{Using Dielectric Material}}},\ }\href
  {https://doi.org/10.1103/PhysRevLett.114.223901} {\bibfield  {journal}
  {\bibinfo  {journal} {Phys. Rev. Lett.}\ }\textbf {\bibinfo {volume} {114}},\
  \bibinfo {pages} {223901} (\bibinfo {year} {2015})}\BibitemShut {NoStop}%
\bibitem [{\citenamefont {Dong}\ \emph {et~al.}(2017)\citenamefont {Dong},
  \citenamefont {Chen}, \citenamefont {Zhu}, \citenamefont {Wang},\ and\
  \citenamefont {Zhang}}]{Dong_Valley_2017}%
  \BibitemOpen
  \bibfield  {author} {\bibinfo {author} {\bibfnamefont {J.-W.}\ \bibnamefont
  {Dong}}, \bibinfo {author} {\bibfnamefont {X.-D.}\ \bibnamefont {Chen}},
  \bibinfo {author} {\bibfnamefont {H.}~\bibnamefont {Zhu}}, \bibinfo {author}
  {\bibfnamefont {Y.}~\bibnamefont {Wang}},\ and\ \bibinfo {author}
  {\bibfnamefont {X.}~\bibnamefont {Zhang}},\ }\bibfield  {title} {\bibinfo
  {title} {Valley photonic crystals for control of spin and topology},\ }\href
  {https://doi.org/10.1038/nmat4807} {\bibfield  {journal} {\bibinfo  {journal}
  {Nat. Mater.}\ }\textbf {\bibinfo {volume} {16}},\ \bibinfo {pages} {298}
  (\bibinfo {year} {2017})}\BibitemShut {NoStop}%
\bibitem [{\citenamefont {Shalaev}\ \emph {et~al.}(2019)\citenamefont
  {Shalaev}, \citenamefont {Walasik}, \citenamefont {Tsukernik}, \citenamefont
  {Xu},\ and\ \citenamefont {Litchinitser}}]{Shalaev_Robust_2019}%
  \BibitemOpen
  \bibfield  {author} {\bibinfo {author} {\bibfnamefont {M.~I.}\ \bibnamefont
  {Shalaev}}, \bibinfo {author} {\bibfnamefont {W.}~\bibnamefont {Walasik}},
  \bibinfo {author} {\bibfnamefont {A.}~\bibnamefont {Tsukernik}}, \bibinfo
  {author} {\bibfnamefont {Y.}~\bibnamefont {Xu}},\ and\ \bibinfo {author}
  {\bibfnamefont {N.~M.}\ \bibnamefont {Litchinitser}},\ }\bibfield  {title}
  {\bibinfo {title} {Robust topologically protected transport in photonic
  crystals at telecommunication wavelengths},\ }\href
  {https://doi.org/10.1038/s41565-018-0297-6} {\bibfield  {journal} {\bibinfo
  {journal} {Nat. Nanotechnol.}\ }\textbf {\bibinfo {volume} {14}},\ \bibinfo
  {pages} {31} (\bibinfo {year} {2019})}\BibitemShut {NoStop}%
\bibitem [{\citenamefont {Liu}\ \emph {et~al.}(2020)\citenamefont {Liu},
  \citenamefont {Ji}, \citenamefont {Wang}, \citenamefont {Modi}, \citenamefont
  {Hwang}, \citenamefont {Zheng}, \citenamefont {Sorger}, \citenamefont {Pan},\
  and\ \citenamefont {Agarwal}}]{Liu_Generation_2020b}%
  \BibitemOpen
  \bibfield  {author} {\bibinfo {author} {\bibfnamefont {W.}~\bibnamefont
  {Liu}}, \bibinfo {author} {\bibfnamefont {Z.}~\bibnamefont {Ji}}, \bibinfo
  {author} {\bibfnamefont {Y.}~\bibnamefont {Wang}}, \bibinfo {author}
  {\bibfnamefont {G.}~\bibnamefont {Modi}}, \bibinfo {author} {\bibfnamefont
  {M.}~\bibnamefont {Hwang}}, \bibinfo {author} {\bibfnamefont
  {B.}~\bibnamefont {Zheng}}, \bibinfo {author} {\bibfnamefont {V.~J.}\
  \bibnamefont {Sorger}}, \bibinfo {author} {\bibfnamefont {A.}~\bibnamefont
  {Pan}},\ and\ \bibinfo {author} {\bibfnamefont {R.}~\bibnamefont {Agarwal}},\
  }\bibfield  {title} {\bibinfo {title} {Generation of helical topological
  exciton-polaritons},\ }\href {https://doi.org/10.1126/science.abc4975}
  {\bibfield  {journal} {\bibinfo  {journal} {Science}\ }\textbf {\bibinfo
  {volume} {370}},\ \bibinfo {pages} {600} (\bibinfo {year}
  {2020})}\BibitemShut {NoStop}%
\bibitem [{\citenamefont {Li}\ \emph {et~al.}(2021)\citenamefont {Li},
  \citenamefont {Sinev}, \citenamefont {Benimetskiy}, \citenamefont {Ivanova},
  \citenamefont {Khestanova}, \citenamefont {Kiriushechkina}, \citenamefont
  {Vakulenko}, \citenamefont {Guddala}, \citenamefont {Skolnick}, \citenamefont
  {Menon}, \citenamefont {Krizhanovskii}, \citenamefont {Al{\`u}},
  \citenamefont {Samusev},\ and\ \citenamefont
  {Khanikaev}}]{Li_Experimental_2021a}%
  \BibitemOpen
  \bibfield  {author} {\bibinfo {author} {\bibfnamefont {M.}~\bibnamefont
  {Li}}, \bibinfo {author} {\bibfnamefont {I.}~\bibnamefont {Sinev}}, \bibinfo
  {author} {\bibfnamefont {F.}~\bibnamefont {Benimetskiy}}, \bibinfo {author}
  {\bibfnamefont {T.}~\bibnamefont {Ivanova}}, \bibinfo {author} {\bibfnamefont
  {E.}~\bibnamefont {Khestanova}}, \bibinfo {author} {\bibfnamefont
  {S.}~\bibnamefont {Kiriushechkina}}, \bibinfo {author} {\bibfnamefont
  {A.}~\bibnamefont {Vakulenko}}, \bibinfo {author} {\bibfnamefont
  {S.}~\bibnamefont {Guddala}}, \bibinfo {author} {\bibfnamefont
  {M.}~\bibnamefont {Skolnick}}, \bibinfo {author} {\bibfnamefont {V.~M.}\
  \bibnamefont {Menon}}, \bibinfo {author} {\bibfnamefont {D.}~\bibnamefont
  {Krizhanovskii}}, \bibinfo {author} {\bibfnamefont {A.}~\bibnamefont
  {Al{\`u}}}, \bibinfo {author} {\bibfnamefont {A.}~\bibnamefont {Samusev}},\
  and\ \bibinfo {author} {\bibfnamefont {A.~B.}\ \bibnamefont {Khanikaev}},\
  }\bibfield  {title} {\bibinfo {title} {Experimental observation of
  topological {{Z2}} exciton-polaritons in transition metal dichalcogenide
  monolayers},\ }\href {https://doi.org/10.1038/s41467-021-24728-y} {\bibfield
  {journal} {\bibinfo  {journal} {Nat. Commun.}\ }\textbf {\bibinfo {volume}
  {12}},\ \bibinfo {pages} {4425} (\bibinfo {year} {2021})}\BibitemShut
  {NoStop}%
\bibitem [{\citenamefont {Arregui}\ \emph {et~al.}(2021)\citenamefont
  {Arregui}, \citenamefont {{Gomis-Bresco}}, \citenamefont
  {{Sotomayor-Torres}},\ and\ \citenamefont
  {Garcia}}]{Arregui_Quantifying_2021}%
  \BibitemOpen
  \bibfield  {author} {\bibinfo {author} {\bibfnamefont {G.}~\bibnamefont
  {Arregui}}, \bibinfo {author} {\bibfnamefont {J.}~\bibnamefont
  {{Gomis-Bresco}}}, \bibinfo {author} {\bibfnamefont {C.~M.}\ \bibnamefont
  {{Sotomayor-Torres}}},\ and\ \bibinfo {author} {\bibfnamefont {P.~D.}\
  \bibnamefont {Garcia}},\ }\bibfield  {title} {\bibinfo {title} {Quantifying
  the {{Robustness}} of {{Topological Slow Light}}},\ }\href
  {https://doi.org/10.1103/PhysRevLett.126.027403} {\bibfield  {journal}
  {\bibinfo  {journal} {Phys. Rev. Lett.}\ }\textbf {\bibinfo {volume} {126}},\
  \bibinfo {pages} {027403} (\bibinfo {year} {2021})}\BibitemShut {NoStop}%
\bibitem [{\citenamefont {Rosiek}\ \emph {et~al.}(2023)\citenamefont {Rosiek},
  \citenamefont {Arregui}, \citenamefont {Vladimirova}, \citenamefont
  {Albrechtsen}, \citenamefont {Vosoughi~Lahijani}, \citenamefont
  {Christiansen},\ and\ \citenamefont {Stobbe}}]{Rosiek_Observation_2023}%
  \BibitemOpen
  \bibfield  {author} {\bibinfo {author} {\bibfnamefont {C.~A.}\ \bibnamefont
  {Rosiek}}, \bibinfo {author} {\bibfnamefont {G.}~\bibnamefont {Arregui}},
  \bibinfo {author} {\bibfnamefont {A.}~\bibnamefont {Vladimirova}}, \bibinfo
  {author} {\bibfnamefont {M.}~\bibnamefont {Albrechtsen}}, \bibinfo {author}
  {\bibfnamefont {B.}~\bibnamefont {Vosoughi~Lahijani}}, \bibinfo {author}
  {\bibfnamefont {R.~E.}\ \bibnamefont {Christiansen}},\ and\ \bibinfo {author}
  {\bibfnamefont {S.}~\bibnamefont {Stobbe}},\ }\bibfield  {title} {\bibinfo
  {title} {Observation of strong backscattering in valley-{{Hall}} photonic
  topological interface modes},\ }\href
  {https://doi.org/10.1038/s41566-023-01189-x} {\bibfield  {journal} {\bibinfo
  {journal} {Nat. Photonics}\ }\textbf {\bibinfo {volume} {17}},\ \bibinfo
  {pages} {386} (\bibinfo {year} {2023})}\BibitemShut {NoStop}%
\bibitem [{\citenamefont {Khanikaev}\ and\ \citenamefont
  {Al{\`u}}(2024)}]{Khanikaev_Topological_2024}%
  \BibitemOpen
  \bibfield  {author} {\bibinfo {author} {\bibfnamefont {A.~B.}\ \bibnamefont
  {Khanikaev}}\ and\ \bibinfo {author} {\bibfnamefont {A.}~\bibnamefont
  {Al{\`u}}},\ }\bibfield  {title} {\bibinfo {title} {Topological photonics:
  Robustness and beyond},\ }\href {https://doi.org/10.1038/s41467-024-45194-2}
  {\bibfield  {journal} {\bibinfo  {journal} {Nat. Commun.}\ }\textbf {\bibinfo
  {volume} {15}},\ \bibinfo {pages} {931} (\bibinfo {year} {2024})}\BibitemShut
  {NoStop}%
\bibitem [{\citenamefont {Haldane}\ and\ \citenamefont
  {Raghu}(2008)}]{Haldane_Possible_2008}%
  \BibitemOpen
  \bibfield  {author} {\bibinfo {author} {\bibfnamefont {F.~D.~M.}\
  \bibnamefont {Haldane}}\ and\ \bibinfo {author} {\bibfnamefont
  {S.}~\bibnamefont {Raghu}},\ }\bibfield  {title} {\bibinfo {title} {Possible
  {{Realization}} of {{Directional Optical Waveguides}} in {{Photonic
  Crystals}} with {{Broken Time-Reversal Symmetry}}},\ }\href
  {https://doi.org/10.1103/PhysRevLett.100.013904} {\bibfield  {journal}
  {\bibinfo  {journal} {Phys. Rev. Lett.}\ }\textbf {\bibinfo {volume} {100}},\
  \bibinfo {pages} {013904} (\bibinfo {year} {2008})}\BibitemShut {NoStop}%
\bibitem [{\citenamefont {Raghu}\ and\ \citenamefont
  {Haldane}(2008)}]{Raghu_Analogs_2008}%
  \BibitemOpen
  \bibfield  {author} {\bibinfo {author} {\bibfnamefont {S.}~\bibnamefont
  {Raghu}}\ and\ \bibinfo {author} {\bibfnamefont {F.~D.~M.}\ \bibnamefont
  {Haldane}},\ }\bibfield  {title} {\bibinfo {title} {Analogs of
  quantum-{{Hall-effect}} edge states in photonic crystals},\ }\href
  {https://doi.org/10.1103/PhysRevA.78.033834} {\bibfield  {journal} {\bibinfo
  {journal} {Phys. Rev. A}\ }\textbf {\bibinfo {volume} {78}},\ \bibinfo
  {pages} {033834} (\bibinfo {year} {2008})}\BibitemShut {NoStop}%
\bibitem [{\citenamefont {Wang}\ \emph {et~al.}(2009)\citenamefont {Wang},
  \citenamefont {Chong}, \citenamefont {Joannopoulos},\ and\ \citenamefont
  {Solja{\v c}i{\'c}}}]{Wang_Observation_2009}%
  \BibitemOpen
  \bibfield  {author} {\bibinfo {author} {\bibfnamefont {Z.}~\bibnamefont
  {Wang}}, \bibinfo {author} {\bibfnamefont {Y.}~\bibnamefont {Chong}},
  \bibinfo {author} {\bibfnamefont {J.~D.}\ \bibnamefont {Joannopoulos}},\ and\
  \bibinfo {author} {\bibfnamefont {M.}~\bibnamefont {Solja{\v c}i{\'c}}},\
  }\bibfield  {title} {\bibinfo {title} {Observation of unidirectional
  backscattering-immune topological electromagnetic states},\ }\href
  {https://doi.org/10.1038/nature08293} {\bibfield  {journal} {\bibinfo
  {journal} {Nature}\ }\textbf {\bibinfo {volume} {461}},\ \bibinfo {pages}
  {772} (\bibinfo {year} {2009})}\BibitemShut {NoStop}%
\bibitem [{\citenamefont {Rechtsman}\ \emph {et~al.}(2013)\citenamefont
  {Rechtsman}, \citenamefont {Zeuner}, \citenamefont {Plotnik}, \citenamefont
  {Lumer}, \citenamefont {Podolsky}, \citenamefont {Dreisow}, \citenamefont
  {Nolte}, \citenamefont {Segev},\ and\ \citenamefont
  {Szameit}}]{Rechtsman_Photonic_2013}%
  \BibitemOpen
  \bibfield  {author} {\bibinfo {author} {\bibfnamefont {M.~C.}\ \bibnamefont
  {Rechtsman}}, \bibinfo {author} {\bibfnamefont {J.~M.}\ \bibnamefont
  {Zeuner}}, \bibinfo {author} {\bibfnamefont {Y.}~\bibnamefont {Plotnik}},
  \bibinfo {author} {\bibfnamefont {Y.}~\bibnamefont {Lumer}}, \bibinfo
  {author} {\bibfnamefont {D.}~\bibnamefont {Podolsky}}, \bibinfo {author}
  {\bibfnamefont {F.}~\bibnamefont {Dreisow}}, \bibinfo {author} {\bibfnamefont
  {S.}~\bibnamefont {Nolte}}, \bibinfo {author} {\bibfnamefont
  {M.}~\bibnamefont {Segev}},\ and\ \bibinfo {author} {\bibfnamefont
  {A.}~\bibnamefont {Szameit}},\ }\bibfield  {title} {\bibinfo {title}
  {Photonic {{Floquet}} topological insulators},\ }\href
  {https://doi.org/10.1038/nature12066} {\bibfield  {journal} {\bibinfo
  {journal} {Nature}\ }\textbf {\bibinfo {volume} {496}},\ \bibinfo {pages}
  {196} (\bibinfo {year} {2013})}\BibitemShut {NoStop}%
\bibitem [{\citenamefont {Lustig}\ \emph {et~al.}(2019)\citenamefont {Lustig},
  \citenamefont {Weimann}, \citenamefont {Plotnik}, \citenamefont {Lumer},
  \citenamefont {Bandres}, \citenamefont {Szameit},\ and\ \citenamefont
  {Segev}}]{Lustig_Photonic_2019}%
  \BibitemOpen
  \bibfield  {author} {\bibinfo {author} {\bibfnamefont {E.}~\bibnamefont
  {Lustig}}, \bibinfo {author} {\bibfnamefont {S.}~\bibnamefont {Weimann}},
  \bibinfo {author} {\bibfnamefont {Y.}~\bibnamefont {Plotnik}}, \bibinfo
  {author} {\bibfnamefont {Y.}~\bibnamefont {Lumer}}, \bibinfo {author}
  {\bibfnamefont {M.~A.}\ \bibnamefont {Bandres}}, \bibinfo {author}
  {\bibfnamefont {A.}~\bibnamefont {Szameit}},\ and\ \bibinfo {author}
  {\bibfnamefont {M.}~\bibnamefont {Segev}},\ }\bibfield  {title} {\bibinfo
  {title} {Photonic topological insulator in synthetic dimensions},\ }\href
  {https://doi.org/10.1038/s41586-019-0943-7} {\bibfield  {journal} {\bibinfo
  {journal} {Nature}\ }\textbf {\bibinfo {volume} {567}},\ \bibinfo {pages}
  {356} (\bibinfo {year} {2019})}\BibitemShut {NoStop}%
\bibitem [{\citenamefont {Lindner}\ \emph {et~al.}(2011)\citenamefont
  {Lindner}, \citenamefont {Refael},\ and\ \citenamefont
  {Galitski}}]{Lindner_Floquet_2011}%
  \BibitemOpen
  \bibfield  {author} {\bibinfo {author} {\bibfnamefont {N.~H.}\ \bibnamefont
  {Lindner}}, \bibinfo {author} {\bibfnamefont {G.}~\bibnamefont {Refael}},\
  and\ \bibinfo {author} {\bibfnamefont {V.}~\bibnamefont {Galitski}},\
  }\bibfield  {title} {\bibinfo {title} {Floquet topological insulator in
  semiconductor quantum wells},\ }\href {https://doi.org/10.1038/nphys1926}
  {\bibfield  {journal} {\bibinfo  {journal} {Nat. Phys.}\ }\textbf {\bibinfo
  {volume} {7}},\ \bibinfo {pages} {490} (\bibinfo {year} {2011})}\BibitemShut
  {NoStop}%
\bibitem [{\citenamefont {Fang}\ \emph {et~al.}(2012)\citenamefont {Fang},
  \citenamefont {Yu},\ and\ \citenamefont {Fan}}]{Fang_Realizing_2012}%
  \BibitemOpen
  \bibfield  {author} {\bibinfo {author} {\bibfnamefont {K.}~\bibnamefont
  {Fang}}, \bibinfo {author} {\bibfnamefont {Z.}~\bibnamefont {Yu}},\ and\
  \bibinfo {author} {\bibfnamefont {S.}~\bibnamefont {Fan}},\ }\bibfield
  {title} {\bibinfo {title} {Realizing effective magnetic field for photons by
  controlling the phase of dynamic modulation},\ }\href
  {https://doi.org/10.1038/nphoton.2012.236} {\bibfield  {journal} {\bibinfo
  {journal} {Nat. Photonics}\ }\textbf {\bibinfo {volume} {6}},\ \bibinfo
  {pages} {782} (\bibinfo {year} {2012})}\BibitemShut {NoStop}%
\bibitem [{\citenamefont {Maczewsky}\ \emph {et~al.}(2017)\citenamefont
  {Maczewsky}, \citenamefont {Zeuner}, \citenamefont {Nolte},\ and\
  \citenamefont {Szameit}}]{Maczewsky_Observation_2017}%
  \BibitemOpen
  \bibfield  {author} {\bibinfo {author} {\bibfnamefont {L.~J.}\ \bibnamefont
  {Maczewsky}}, \bibinfo {author} {\bibfnamefont {J.~M.}\ \bibnamefont
  {Zeuner}}, \bibinfo {author} {\bibfnamefont {S.}~\bibnamefont {Nolte}},\ and\
  \bibinfo {author} {\bibfnamefont {A.}~\bibnamefont {Szameit}},\ }\bibfield
  {title} {\bibinfo {title} {Observation of photonic anomalous {{Floquet}}
  topological insulators},\ }\href {https://doi.org/10.1038/ncomms13756}
  {\bibfield  {journal} {\bibinfo  {journal} {Nat. Commun.}\ }\textbf {\bibinfo
  {volume} {8}},\ \bibinfo {pages} {13756} (\bibinfo {year}
  {2017})}\BibitemShut {NoStop}%
\bibitem [{\citenamefont {Zhao}\ \emph {et~al.}(2019)\citenamefont {Zhao},
  \citenamefont {Qiao}, \citenamefont {Wu}, \citenamefont {Midya},
  \citenamefont {Longhi},\ and\ \citenamefont {Feng}}]{Zhao_NonHermitian_2019}%
  \BibitemOpen
  \bibfield  {author} {\bibinfo {author} {\bibfnamefont {H.}~\bibnamefont
  {Zhao}}, \bibinfo {author} {\bibfnamefont {X.}~\bibnamefont {Qiao}}, \bibinfo
  {author} {\bibfnamefont {T.}~\bibnamefont {Wu}}, \bibinfo {author}
  {\bibfnamefont {B.}~\bibnamefont {Midya}}, \bibinfo {author} {\bibfnamefont
  {S.}~\bibnamefont {Longhi}},\ and\ \bibinfo {author} {\bibfnamefont
  {L.}~\bibnamefont {Feng}},\ }\bibfield  {title} {\bibinfo {title}
  {Non-{{Hermitian}} topological light steering},\ }\href
  {https://doi.org/10.1126/science.aay1064} {\bibfield  {journal} {\bibinfo
  {journal} {Science}\ }\textbf {\bibinfo {volume} {365}},\ \bibinfo {pages}
  {1163} (\bibinfo {year} {2019})}\BibitemShut {NoStop}%
\bibitem [{\citenamefont {Xiao}\ \emph {et~al.}(2020)\citenamefont {Xiao},
  \citenamefont {Deng}, \citenamefont {Wang}, \citenamefont {Zhu},
  \citenamefont {Wang}, \citenamefont {Yi},\ and\ \citenamefont
  {Xue}}]{Xiao_NonHermitian_2020}%
  \BibitemOpen
  \bibfield  {author} {\bibinfo {author} {\bibfnamefont {L.}~\bibnamefont
  {Xiao}}, \bibinfo {author} {\bibfnamefont {T.}~\bibnamefont {Deng}}, \bibinfo
  {author} {\bibfnamefont {K.}~\bibnamefont {Wang}}, \bibinfo {author}
  {\bibfnamefont {G.}~\bibnamefont {Zhu}}, \bibinfo {author} {\bibfnamefont
  {Z.}~\bibnamefont {Wang}}, \bibinfo {author} {\bibfnamefont {W.}~\bibnamefont
  {Yi}},\ and\ \bibinfo {author} {\bibfnamefont {P.}~\bibnamefont {Xue}},\
  }\bibfield  {title} {\bibinfo {title} {Non-{{Hermitian}} bulk--boundary
  correspondence in quantum dynamics},\ }\href
  {https://doi.org/10.1038/s41567-020-0836-6} {\bibfield  {journal} {\bibinfo
  {journal} {Nat. Phys.}\ }\textbf {\bibinfo {volume} {16}},\ \bibinfo {pages}
  {761} (\bibinfo {year} {2020})}\BibitemShut {NoStop}%
\bibitem [{\citenamefont {Dai}\ \emph {et~al.}(2024)\citenamefont {Dai},
  \citenamefont {Ao}, \citenamefont {Mao}, \citenamefont {Yang}, \citenamefont
  {Zheng}, \citenamefont {Zhai}, \citenamefont {Li}, \citenamefont {Yuan},
  \citenamefont {Tang}, \citenamefont {Li}, \citenamefont {Luo}, \citenamefont
  {Wang}, \citenamefont {Hu}, \citenamefont {Gong},\ and\ \citenamefont
  {Wang}}]{Dai_NonHermitian_2024}%
  \BibitemOpen
  \bibfield  {author} {\bibinfo {author} {\bibfnamefont {T.}~\bibnamefont
  {Dai}}, \bibinfo {author} {\bibfnamefont {Y.}~\bibnamefont {Ao}}, \bibinfo
  {author} {\bibfnamefont {J.}~\bibnamefont {Mao}}, \bibinfo {author}
  {\bibfnamefont {Y.}~\bibnamefont {Yang}}, \bibinfo {author} {\bibfnamefont
  {Y.}~\bibnamefont {Zheng}}, \bibinfo {author} {\bibfnamefont
  {C.}~\bibnamefont {Zhai}}, \bibinfo {author} {\bibfnamefont {Y.}~\bibnamefont
  {Li}}, \bibinfo {author} {\bibfnamefont {J.}~\bibnamefont {Yuan}}, \bibinfo
  {author} {\bibfnamefont {B.}~\bibnamefont {Tang}}, \bibinfo {author}
  {\bibfnamefont {Z.}~\bibnamefont {Li}}, \bibinfo {author} {\bibfnamefont
  {J.}~\bibnamefont {Luo}}, \bibinfo {author} {\bibfnamefont {W.}~\bibnamefont
  {Wang}}, \bibinfo {author} {\bibfnamefont {X.}~\bibnamefont {Hu}}, \bibinfo
  {author} {\bibfnamefont {Q.}~\bibnamefont {Gong}},\ and\ \bibinfo {author}
  {\bibfnamefont {J.}~\bibnamefont {Wang}},\ }\bibfield  {title} {\bibinfo
  {title} {Non-{{Hermitian}} topological phase transitions controlled by
  nonlinearity},\ }\href {https://doi.org/10.1038/s41567-023-02244-8}
  {\bibfield  {journal} {\bibinfo  {journal} {Nat. Phys.}\ }\textbf {\bibinfo
  {volume} {20}},\ \bibinfo {pages} {101} (\bibinfo {year} {2024})}\BibitemShut
  {NoStop}%
\bibitem [{\citenamefont {Weisbuch}\ \emph {et~al.}(1992)\citenamefont
  {Weisbuch}, \citenamefont {Nishioka}, \citenamefont {Ishikawa},\ and\
  \citenamefont {Arakawa}}]{weisbuch_observation_1992}%
  \BibitemOpen
  \bibfield  {author} {\bibinfo {author} {\bibfnamefont {C.}~\bibnamefont
  {Weisbuch}}, \bibinfo {author} {\bibfnamefont {M.}~\bibnamefont {Nishioka}},
  \bibinfo {author} {\bibfnamefont {A.}~\bibnamefont {Ishikawa}},\ and\
  \bibinfo {author} {\bibfnamefont {Y.}~\bibnamefont {Arakawa}},\ }\bibfield
  {title} {\bibinfo {title} {Observation of the coupled exciton-photon mode
  splitting in a semiconductor quantum microcavity},\ }\href
  {https://doi.org/10.1103/PhysRevLett.69.3314} {\bibfield  {journal} {\bibinfo
   {journal} {Phys. Rev. Lett.}\ }\textbf {\bibinfo {volume} {69}},\ \bibinfo
  {pages} {3314} (\bibinfo {year} {1992})}\BibitemShut {NoStop}%
\bibitem [{\citenamefont {Karzig}\ \emph {et~al.}(2015)\citenamefont {Karzig},
  \citenamefont {Bardyn}, \citenamefont {Lindner},\ and\ \citenamefont
  {Refael}}]{Karzig_Topological_2015}%
  \BibitemOpen
  \bibfield  {author} {\bibinfo {author} {\bibfnamefont {T.}~\bibnamefont
  {Karzig}}, \bibinfo {author} {\bibfnamefont {C.-E.}\ \bibnamefont {Bardyn}},
  \bibinfo {author} {\bibfnamefont {N.~H.}\ \bibnamefont {Lindner}},\ and\
  \bibinfo {author} {\bibfnamefont {G.}~\bibnamefont {Refael}},\ }\bibfield
  {title} {\bibinfo {title} {Topological {{Polaritons}}},\ }\href
  {https://doi.org/10.1103/PhysRevX.5.031001} {\bibfield  {journal} {\bibinfo
  {journal} {Phys. Rev. X}\ }\textbf {\bibinfo {volume} {5}},\ \bibinfo {pages}
  {031001} (\bibinfo {year} {2015})}\BibitemShut {NoStop}%
\bibitem [{\citenamefont {Gianfrate}\ \emph {et~al.}(2020)\citenamefont
  {Gianfrate}, \citenamefont {Bleu}, \citenamefont {Dominici}, \citenamefont
  {Ardizzone}, \citenamefont {De~Giorgi}, \citenamefont {Ballarini},
  \citenamefont {Lerario}, \citenamefont {West}, \citenamefont {Pfeiffer},
  \citenamefont {Solnyshkov}, \citenamefont {Sanvitto},\ and\ \citenamefont
  {Malpuech}}]{Gianfrate_Measurement_2020}%
  \BibitemOpen
  \bibfield  {author} {\bibinfo {author} {\bibfnamefont {A.}~\bibnamefont
  {Gianfrate}}, \bibinfo {author} {\bibfnamefont {O.}~\bibnamefont {Bleu}},
  \bibinfo {author} {\bibfnamefont {L.}~\bibnamefont {Dominici}}, \bibinfo
  {author} {\bibfnamefont {V.}~\bibnamefont {Ardizzone}}, \bibinfo {author}
  {\bibfnamefont {M.}~\bibnamefont {De~Giorgi}}, \bibinfo {author}
  {\bibfnamefont {D.}~\bibnamefont {Ballarini}}, \bibinfo {author}
  {\bibfnamefont {G.}~\bibnamefont {Lerario}}, \bibinfo {author} {\bibfnamefont
  {K.~W.}\ \bibnamefont {West}}, \bibinfo {author} {\bibfnamefont {L.~N.}\
  \bibnamefont {Pfeiffer}}, \bibinfo {author} {\bibfnamefont {D.~D.}\
  \bibnamefont {Solnyshkov}}, \bibinfo {author} {\bibfnamefont
  {D.}~\bibnamefont {Sanvitto}},\ and\ \bibinfo {author} {\bibfnamefont
  {G.}~\bibnamefont {Malpuech}},\ }\bibfield  {title} {\bibinfo {title}
  {Measurement of the quantum geometric tensor and of the anomalous {{Hall}}
  drift},\ }\href {https://doi.org/10.1038/s41586-020-1989-2} {\bibfield
  {journal} {\bibinfo  {journal} {Nature}\ }\textbf {\bibinfo {volume} {578}},\
  \bibinfo {pages} {381} (\bibinfo {year} {2020})}\BibitemShut {NoStop}%
\bibitem [{\citenamefont {Polimeno}\ \emph {et~al.}(2021)\citenamefont
  {Polimeno}, \citenamefont {Lerario}, \citenamefont {De~Giorgi}, \citenamefont
  {De~Marco}, \citenamefont {Dominici}, \citenamefont {Todisco}, \citenamefont
  {Coriolano}, \citenamefont {Ardizzone}, \citenamefont {Pugliese},
  \citenamefont {Prontera}, \citenamefont {Maiorano}, \citenamefont
  {Moliterni}, \citenamefont {Giannini}, \citenamefont {Olieric}, \citenamefont
  {Gigli}, \citenamefont {Ballarini}, \citenamefont {Xiong}, \citenamefont
  {Fieramosca}, \citenamefont {Solnyshkov}, \citenamefont {Malpuech},\ and\
  \citenamefont {Sanvitto}}]{Polimeno_Tuning_2021}%
  \BibitemOpen
  \bibfield  {author} {\bibinfo {author} {\bibfnamefont {L.}~\bibnamefont
  {Polimeno}}, \bibinfo {author} {\bibfnamefont {G.}~\bibnamefont {Lerario}},
  \bibinfo {author} {\bibfnamefont {M.}~\bibnamefont {De~Giorgi}}, \bibinfo
  {author} {\bibfnamefont {L.}~\bibnamefont {De~Marco}}, \bibinfo {author}
  {\bibfnamefont {L.}~\bibnamefont {Dominici}}, \bibinfo {author}
  {\bibfnamefont {F.}~\bibnamefont {Todisco}}, \bibinfo {author} {\bibfnamefont
  {A.}~\bibnamefont {Coriolano}}, \bibinfo {author} {\bibfnamefont
  {V.}~\bibnamefont {Ardizzone}}, \bibinfo {author} {\bibfnamefont
  {M.}~\bibnamefont {Pugliese}}, \bibinfo {author} {\bibfnamefont {C.~T.}\
  \bibnamefont {Prontera}}, \bibinfo {author} {\bibfnamefont {V.}~\bibnamefont
  {Maiorano}}, \bibinfo {author} {\bibfnamefont {A.}~\bibnamefont {Moliterni}},
  \bibinfo {author} {\bibfnamefont {C.}~\bibnamefont {Giannini}}, \bibinfo
  {author} {\bibfnamefont {V.}~\bibnamefont {Olieric}}, \bibinfo {author}
  {\bibfnamefont {G.}~\bibnamefont {Gigli}}, \bibinfo {author} {\bibfnamefont
  {D.}~\bibnamefont {Ballarini}}, \bibinfo {author} {\bibfnamefont
  {Q.}~\bibnamefont {Xiong}}, \bibinfo {author} {\bibfnamefont
  {A.}~\bibnamefont {Fieramosca}}, \bibinfo {author} {\bibfnamefont {D.~D.}\
  \bibnamefont {Solnyshkov}}, \bibinfo {author} {\bibfnamefont
  {G.}~\bibnamefont {Malpuech}},\ and\ \bibinfo {author} {\bibfnamefont
  {D.}~\bibnamefont {Sanvitto}},\ }\bibfield  {title} {\bibinfo {title} {Tuning
  of the {{Berry}} curvature in {{2D}} perovskite polaritons},\ }\href
  {https://doi.org/10.1038/s41565-021-00977-2} {\bibfield  {journal} {\bibinfo
  {journal} {Nat. Nanotechnol.}\ }\textbf {\bibinfo {volume} {16}},\ \bibinfo
  {pages} {1349} (\bibinfo {year} {2021})}\BibitemShut {NoStop}%
\bibitem [{\citenamefont {Nalitov}\ \emph {et~al.}(2015)\citenamefont
  {Nalitov}, \citenamefont {Solnyshkov},\ and\ \citenamefont
  {Malpuech}}]{Nalitov_Polariton_2015}%
  \BibitemOpen
  \bibfield  {author} {\bibinfo {author} {\bibfnamefont {A.~V.}\ \bibnamefont
  {Nalitov}}, \bibinfo {author} {\bibfnamefont {D.~D.}\ \bibnamefont
  {Solnyshkov}},\ and\ \bibinfo {author} {\bibfnamefont {G.}~\bibnamefont
  {Malpuech}},\ }\bibfield  {title} {\bibinfo {title} {Polariton $\mathbb{Z}$
  {{Topological Insulator}}},\ }\href
  {https://doi.org/10.1103/PhysRevLett.114.116401} {\bibfield  {journal}
  {\bibinfo  {journal} {Phys. Rev. Lett.}\ }\textbf {\bibinfo {volume} {114}},\
  \bibinfo {pages} {116401} (\bibinfo {year} {2015})}\BibitemShut {NoStop}%
\bibitem [{\citenamefont {Bardyn}\ \emph {et~al.}(2015)\citenamefont {Bardyn},
  \citenamefont {Karzig}, \citenamefont {Refael},\ and\ \citenamefont
  {Liew}}]{Bardyn_Topological_2015}%
  \BibitemOpen
  \bibfield  {author} {\bibinfo {author} {\bibfnamefont {C.-E.}\ \bibnamefont
  {Bardyn}}, \bibinfo {author} {\bibfnamefont {T.}~\bibnamefont {Karzig}},
  \bibinfo {author} {\bibfnamefont {G.}~\bibnamefont {Refael}},\ and\ \bibinfo
  {author} {\bibfnamefont {T.~C.~H.}\ \bibnamefont {Liew}},\ }\bibfield
  {title} {\bibinfo {title} {Topological polaritons and excitons in
  garden-variety systems},\ }\href {https://doi.org/10.1103/PhysRevB.91.161413}
  {\bibfield  {journal} {\bibinfo  {journal} {Phys. Rev. B}\ }\textbf {\bibinfo
  {volume} {91}},\ \bibinfo {pages} {161413} (\bibinfo {year}
  {2015})}\BibitemShut {NoStop}%
\bibitem [{\citenamefont {Yi}\ and\ \citenamefont
  {Karzig}(2016)}]{Yi_Topological_2016}%
  \BibitemOpen
  \bibfield  {author} {\bibinfo {author} {\bibfnamefont {K.}~\bibnamefont
  {Yi}}\ and\ \bibinfo {author} {\bibfnamefont {T.}~\bibnamefont {Karzig}},\
  }\bibfield  {title} {\bibinfo {title} {Topological polaritons from photonic
  {{Dirac}} cones coupled to excitons in a magnetic field},\ }\href
  {https://doi.org/10.1103/PhysRevB.93.104303} {\bibfield  {journal} {\bibinfo
  {journal} {Phys. Rev. B}\ }\textbf {\bibinfo {volume} {93}},\ \bibinfo
  {pages} {104303} (\bibinfo {year} {2016})}\BibitemShut {NoStop}%
\bibitem [{\citenamefont {He}\ \emph {et~al.}(2023)\citenamefont {He},
  \citenamefont {Wu}, \citenamefont {Jin}, \citenamefont {Mele},\ and\
  \citenamefont {Zhen}}]{He_Polaritonic_2023}%
  \BibitemOpen
  \bibfield  {author} {\bibinfo {author} {\bibfnamefont {L.}~\bibnamefont
  {He}}, \bibinfo {author} {\bibfnamefont {J.}~\bibnamefont {Wu}}, \bibinfo
  {author} {\bibfnamefont {J.}~\bibnamefont {Jin}}, \bibinfo {author}
  {\bibfnamefont {E.~J.}\ \bibnamefont {Mele}},\ and\ \bibinfo {author}
  {\bibfnamefont {B.}~\bibnamefont {Zhen}},\ }\bibfield  {title} {\bibinfo
  {title} {Polaritonic {{Chern Insulators}} in {{Monolayer Semiconductors}}},\
  }\href {https://doi.org/10.1103/PhysRevLett.130.043801} {\bibfield  {journal}
  {\bibinfo  {journal} {Phys. Rev. Lett.}\ }\textbf {\bibinfo {volume} {130}},\
  \bibinfo {pages} {043801} (\bibinfo {year} {2023})}\BibitemShut {NoStop}%
\bibitem [{\citenamefont {Klembt}\ \emph {et~al.}(2018)\citenamefont {Klembt},
  \citenamefont {Harder}, \citenamefont {Egorov}, \citenamefont {Winkler},
  \citenamefont {Ge}, \citenamefont {Bandres}, \citenamefont {Emmerling},
  \citenamefont {Worschech}, \citenamefont {Liew}, \citenamefont {Segev},
  \citenamefont {Schneider},\ and\ \citenamefont
  {H{\"o}fling}}]{Klembt_Excitonpolariton_2018a}%
  \BibitemOpen
  \bibfield  {author} {\bibinfo {author} {\bibfnamefont {S.}~\bibnamefont
  {Klembt}}, \bibinfo {author} {\bibfnamefont {T.~H.}\ \bibnamefont {Harder}},
  \bibinfo {author} {\bibfnamefont {O.~A.}\ \bibnamefont {Egorov}}, \bibinfo
  {author} {\bibfnamefont {K.}~\bibnamefont {Winkler}}, \bibinfo {author}
  {\bibfnamefont {R.}~\bibnamefont {Ge}}, \bibinfo {author} {\bibfnamefont
  {M.~A.}\ \bibnamefont {Bandres}}, \bibinfo {author} {\bibfnamefont
  {M.}~\bibnamefont {Emmerling}}, \bibinfo {author} {\bibfnamefont
  {L.}~\bibnamefont {Worschech}}, \bibinfo {author} {\bibfnamefont {T.~C.~H.}\
  \bibnamefont {Liew}}, \bibinfo {author} {\bibfnamefont {M.}~\bibnamefont
  {Segev}}, \bibinfo {author} {\bibfnamefont {C.}~\bibnamefont {Schneider}},\
  and\ \bibinfo {author} {\bibfnamefont {S.}~\bibnamefont {H{\"o}fling}},\
  }\bibfield  {title} {\bibinfo {title} {Exciton-polariton topological
  insulator},\ }\href {https://doi.org/10.1038/s41586-018-0601-5} {\bibfield
  {journal} {\bibinfo  {journal} {Nature}\ }\textbf {\bibinfo {volume} {562}},\
  \bibinfo {pages} {552} (\bibinfo {year} {2018})}\BibitemShut {NoStop}%
\bibitem [{\citenamefont {{von Neumann}}\ and\ \citenamefont
  {Wigner}(1993)}]{vonNeumann_Ueber_1993}%
  \BibitemOpen
  \bibfield  {author} {\bibinfo {author} {\bibfnamefont {J.}~\bibnamefont {{von
  Neumann}}}\ and\ \bibinfo {author} {\bibfnamefont {E.~P.}\ \bibnamefont
  {Wigner}},\ }\bibfield  {title} {\bibinfo {title} {{{\"U}ber merkw{\"u}rdige
  diskrete Eigenwerte}},\ }in\ \href
  {https://doi.org/10.1007/978-3-662-02781-3_19} {\emph {\bibinfo {booktitle}
  {The Collected Works of Eugene Paul Wigner: Part A}}},\ \bibinfo {editor}
  {edited by\ \bibinfo {editor} {\bibfnamefont {A.~S.}\ \bibnamefont
  {Wightman}}}\ (\bibinfo  {publisher} {Springer},\ \bibinfo {address} {Berlin,
  Heidelberg},\ \bibinfo {year} {1993})\ pp.\ \bibinfo {pages}
  {291--293}\BibitemShut {NoStop}%
\bibitem [{\citenamefont {Hsu}\ \emph {et~al.}(2016)\citenamefont {Hsu},
  \citenamefont {Zhen}, \citenamefont {Stone}, \citenamefont {Joannopoulos},\
  and\ \citenamefont {Solja{\v c}i{\'c}}}]{Hsu_Bound_2016}%
  \BibitemOpen
  \bibfield  {author} {\bibinfo {author} {\bibfnamefont {C.~W.}\ \bibnamefont
  {Hsu}}, \bibinfo {author} {\bibfnamefont {B.}~\bibnamefont {Zhen}}, \bibinfo
  {author} {\bibfnamefont {A.~D.}\ \bibnamefont {Stone}}, \bibinfo {author}
  {\bibfnamefont {J.~D.}\ \bibnamefont {Joannopoulos}},\ and\ \bibinfo {author}
  {\bibfnamefont {M.}~\bibnamefont {Solja{\v c}i{\'c}}},\ }\bibfield  {title}
  {\bibinfo {title} {Bound states in the continuum},\ }\href
  {https://doi.org/10.1038/natrevmats.2016.48} {\bibfield  {journal} {\bibinfo
  {journal} {Nat. Rev. Mater.}\ }\textbf {\bibinfo {volume} {1}},\ \bibinfo
  {pages} {1} (\bibinfo {year} {2016})}\BibitemShut {NoStop}%
\bibitem [{\citenamefont {Kang}\ \emph {et~al.}(2023)\citenamefont {Kang},
  \citenamefont {Liu}, \citenamefont {Chan},\ and\ \citenamefont
  {Xiao}}]{Kang_Applications_2023}%
  \BibitemOpen
  \bibfield  {author} {\bibinfo {author} {\bibfnamefont {M.}~\bibnamefont
  {Kang}}, \bibinfo {author} {\bibfnamefont {T.}~\bibnamefont {Liu}}, \bibinfo
  {author} {\bibfnamefont {C.~T.}\ \bibnamefont {Chan}},\ and\ \bibinfo
  {author} {\bibfnamefont {M.}~\bibnamefont {Xiao}},\ }\bibfield  {title}
  {\bibinfo {title} {Applications of bound states in the continuum in
  photonics},\ }\href {https://doi.org/10.1038/s42254-023-00642-8} {\bibfield
  {journal} {\bibinfo  {journal} {Nat. Rev. Phys.}\ }\textbf {\bibinfo {volume}
  {5}},\ \bibinfo {pages} {659} (\bibinfo {year} {2023})}\BibitemShut {NoStop}%
\bibitem [{\citenamefont {Hsu}\ \emph {et~al.}(2013)\citenamefont {Hsu},
  \citenamefont {Zhen}, \citenamefont {Lee}, \citenamefont {Chua},
  \citenamefont {Johnson}, \citenamefont {Joannopoulos},\ and\ \citenamefont
  {Solja{\v c}i{\'c}}}]{Hsu_Observation_2013}%
  \BibitemOpen
  \bibfield  {author} {\bibinfo {author} {\bibfnamefont {C.~W.}\ \bibnamefont
  {Hsu}}, \bibinfo {author} {\bibfnamefont {B.}~\bibnamefont {Zhen}}, \bibinfo
  {author} {\bibfnamefont {J.}~\bibnamefont {Lee}}, \bibinfo {author}
  {\bibfnamefont {S.-L.}\ \bibnamefont {Chua}}, \bibinfo {author}
  {\bibfnamefont {S.~G.}\ \bibnamefont {Johnson}}, \bibinfo {author}
  {\bibfnamefont {J.~D.}\ \bibnamefont {Joannopoulos}},\ and\ \bibinfo {author}
  {\bibfnamefont {M.}~\bibnamefont {Solja{\v c}i{\'c}}},\ }\bibfield  {title}
  {\bibinfo {title} {Observation of trapped light within the radiation
  continuum},\ }\href {https://doi.org/10.1038/nature12289} {\bibfield
  {journal} {\bibinfo  {journal} {Nature}\ }\textbf {\bibinfo {volume} {499}},\
  \bibinfo {pages} {188} (\bibinfo {year} {2013})}\BibitemShut {NoStop}%
\bibitem [{\citenamefont {Jin}\ \emph {et~al.}(2019)\citenamefont {Jin},
  \citenamefont {Yin}, \citenamefont {Ni}, \citenamefont {Solja{\v c}i{\'c}},
  \citenamefont {Zhen},\ and\ \citenamefont {Peng}}]{Jin_Topologically_2019}%
  \BibitemOpen
  \bibfield  {author} {\bibinfo {author} {\bibfnamefont {J.}~\bibnamefont
  {Jin}}, \bibinfo {author} {\bibfnamefont {X.}~\bibnamefont {Yin}}, \bibinfo
  {author} {\bibfnamefont {L.}~\bibnamefont {Ni}}, \bibinfo {author}
  {\bibfnamefont {M.}~\bibnamefont {Solja{\v c}i{\'c}}}, \bibinfo {author}
  {\bibfnamefont {B.}~\bibnamefont {Zhen}},\ and\ \bibinfo {author}
  {\bibfnamefont {C.}~\bibnamefont {Peng}},\ }\bibfield  {title} {\bibinfo
  {title} {Topologically enabled ultrahigh-{{Q}} guided resonances robust to
  out-of-plane scattering},\ }\href {https://doi.org/10.1038/s41586-019-1664-7}
  {\bibfield  {journal} {\bibinfo  {journal} {Nature}\ }\textbf {\bibinfo
  {volume} {574}},\ \bibinfo {pages} {501} (\bibinfo {year}
  {2019})}\BibitemShut {NoStop}%
\bibitem [{\citenamefont {Kang}\ \emph {et~al.}(2022)\citenamefont {Kang},
  \citenamefont {Mao}, \citenamefont {Zhang}, \citenamefont {Xiao},
  \citenamefont {Xu},\ and\ \citenamefont {Chan}}]{Kang_Merging_2022}%
  \BibitemOpen
  \bibfield  {author} {\bibinfo {author} {\bibfnamefont {M.}~\bibnamefont
  {Kang}}, \bibinfo {author} {\bibfnamefont {L.}~\bibnamefont {Mao}}, \bibinfo
  {author} {\bibfnamefont {S.}~\bibnamefont {Zhang}}, \bibinfo {author}
  {\bibfnamefont {M.}~\bibnamefont {Xiao}}, \bibinfo {author} {\bibfnamefont
  {H.}~\bibnamefont {Xu}},\ and\ \bibinfo {author} {\bibfnamefont {C.~T.}\
  \bibnamefont {Chan}},\ }\bibfield  {title} {\bibinfo {title} {Merging bound
  states in the continuum by harnessing higher-order topological charges},\
  }\href {https://doi.org/10.1038/s41377-022-00923-4} {\bibfield  {journal}
  {\bibinfo  {journal} {Light Sci. Appl.}\ }\textbf {\bibinfo {volume} {11}},\
  \bibinfo {pages} {228} (\bibinfo {year} {2022})}\BibitemShut {NoStop}%
\bibitem [{\citenamefont {Zhen}\ \emph {et~al.}(2014)\citenamefont {Zhen},
  \citenamefont {Hsu}, \citenamefont {Lu}, \citenamefont {Stone},\ and\
  \citenamefont {Solja{\v c}i{\'c}}}]{Zhen_Topological_2014}%
  \BibitemOpen
  \bibfield  {author} {\bibinfo {author} {\bibfnamefont {B.}~\bibnamefont
  {Zhen}}, \bibinfo {author} {\bibfnamefont {C.~W.}\ \bibnamefont {Hsu}},
  \bibinfo {author} {\bibfnamefont {L.}~\bibnamefont {Lu}}, \bibinfo {author}
  {\bibfnamefont {A.~D.}\ \bibnamefont {Stone}},\ and\ \bibinfo {author}
  {\bibfnamefont {M.}~\bibnamefont {Solja{\v c}i{\'c}}},\ }\bibfield  {title}
  {\bibinfo {title} {Topological {{Nature}} of {{Optical Bound States}} in the
  {{Continuum}}},\ }\href {https://doi.org/10.1103/PhysRevLett.113.257401}
  {\bibfield  {journal} {\bibinfo  {journal} {Phys. Rev. Lett.}\ }\textbf
  {\bibinfo {volume} {113}},\ \bibinfo {pages} {257401} (\bibinfo {year}
  {2014})}\BibitemShut {NoStop}%
\bibitem [{\citenamefont {Zhang}\ \emph
  {et~al.}(2018{\natexlab{a}})\citenamefont {Zhang}, \citenamefont {Chen},
  \citenamefont {Liu}, \citenamefont {Hsu}, \citenamefont {Wang}, \citenamefont
  {Guan}, \citenamefont {Liu}, \citenamefont {Shi}, \citenamefont {Lu},\ and\
  \citenamefont {Zi}}]{Zhang_Observation_2018}%
  \BibitemOpen
  \bibfield  {author} {\bibinfo {author} {\bibfnamefont {Y.}~\bibnamefont
  {Zhang}}, \bibinfo {author} {\bibfnamefont {A.}~\bibnamefont {Chen}},
  \bibinfo {author} {\bibfnamefont {W.}~\bibnamefont {Liu}}, \bibinfo {author}
  {\bibfnamefont {C.~W.}\ \bibnamefont {Hsu}}, \bibinfo {author} {\bibfnamefont
  {B.}~\bibnamefont {Wang}}, \bibinfo {author} {\bibfnamefont {F.}~\bibnamefont
  {Guan}}, \bibinfo {author} {\bibfnamefont {X.}~\bibnamefont {Liu}}, \bibinfo
  {author} {\bibfnamefont {L.}~\bibnamefont {Shi}}, \bibinfo {author}
  {\bibfnamefont {L.}~\bibnamefont {Lu}},\ and\ \bibinfo {author}
  {\bibfnamefont {J.}~\bibnamefont {Zi}},\ }\bibfield  {title} {\bibinfo
  {title} {Observation of {{Polarization Vortices}} in {{Momentum Space}}},\
  }\href {https://doi.org/10.1103/PhysRevLett.120.186103} {\bibfield  {journal}
  {\bibinfo  {journal} {Phys. Rev. Lett.}\ }\textbf {\bibinfo {volume} {120}},\
  \bibinfo {pages} {186103} (\bibinfo {year} {2018}{\natexlab{a}})}\BibitemShut
  {NoStop}%
\bibitem [{\citenamefont {Yoda}\ and\ \citenamefont
  {Notomi}(2020)}]{Yoda_Generation_2020a}%
  \BibitemOpen
  \bibfield  {author} {\bibinfo {author} {\bibfnamefont {T.}~\bibnamefont
  {Yoda}}\ and\ \bibinfo {author} {\bibfnamefont {M.}~\bibnamefont {Notomi}},\
  }\bibfield  {title} {\bibinfo {title} {Generation and {{Annihilation}} of
  {{Topologically Protected Bound States}} in the {{Continuum}} and
  {{Circularly Polarized States}} by {{Symmetry Breaking}}},\ }\href
  {https://doi.org/10.1103/PhysRevLett.125.053902} {\bibfield  {journal}
  {\bibinfo  {journal} {Phys. Rev. Lett.}\ }\textbf {\bibinfo {volume} {125}},\
  \bibinfo {pages} {053902} (\bibinfo {year} {2020})}\BibitemShut {NoStop}%
\bibitem [{\citenamefont {Wang}\ \emph {et~al.}(2020)\citenamefont {Wang},
  \citenamefont {Liu}, \citenamefont {Zhao}, \citenamefont {Wang},
  \citenamefont {Zhang}, \citenamefont {Chen}, \citenamefont {Guan},
  \citenamefont {Liu}, \citenamefont {Shi},\ and\ \citenamefont
  {Zi}}]{Wang_Generating_2020}%
  \BibitemOpen
  \bibfield  {author} {\bibinfo {author} {\bibfnamefont {B.}~\bibnamefont
  {Wang}}, \bibinfo {author} {\bibfnamefont {W.}~\bibnamefont {Liu}}, \bibinfo
  {author} {\bibfnamefont {M.}~\bibnamefont {Zhao}}, \bibinfo {author}
  {\bibfnamefont {J.}~\bibnamefont {Wang}}, \bibinfo {author} {\bibfnamefont
  {Y.}~\bibnamefont {Zhang}}, \bibinfo {author} {\bibfnamefont
  {A.}~\bibnamefont {Chen}}, \bibinfo {author} {\bibfnamefont {F.}~\bibnamefont
  {Guan}}, \bibinfo {author} {\bibfnamefont {X.}~\bibnamefont {Liu}}, \bibinfo
  {author} {\bibfnamefont {L.}~\bibnamefont {Shi}},\ and\ \bibinfo {author}
  {\bibfnamefont {J.}~\bibnamefont {Zi}},\ }\bibfield  {title} {\bibinfo
  {title} {Generating optical vortex beams by momentum-space polarization
  vortices centred at bound states in the continuum},\ }\href
  {https://doi.org/10.1038/s41566-020-0658-1} {\bibfield  {journal} {\bibinfo
  {journal} {Nat. Photonics}\ }\textbf {\bibinfo {volume} {14}},\ \bibinfo
  {pages} {623} (\bibinfo {year} {2020})}\BibitemShut {NoStop}%
\bibitem [{\citenamefont {Huang}\ \emph {et~al.}(2020)\citenamefont {Huang},
  \citenamefont {Zhang}, \citenamefont {Xiao}, \citenamefont {Wang},
  \citenamefont {Fan}, \citenamefont {Liu}, \citenamefont {Zhang},
  \citenamefont {Qu}, \citenamefont {Ji}, \citenamefont {Han}, \citenamefont
  {Ge}, \citenamefont {Kivshar},\ and\ \citenamefont
  {Song}}]{Huang_Ultrafast_2020}%
  \BibitemOpen
  \bibfield  {author} {\bibinfo {author} {\bibfnamefont {C.}~\bibnamefont
  {Huang}}, \bibinfo {author} {\bibfnamefont {C.}~\bibnamefont {Zhang}},
  \bibinfo {author} {\bibfnamefont {S.}~\bibnamefont {Xiao}}, \bibinfo {author}
  {\bibfnamefont {Y.}~\bibnamefont {Wang}}, \bibinfo {author} {\bibfnamefont
  {Y.}~\bibnamefont {Fan}}, \bibinfo {author} {\bibfnamefont {Y.}~\bibnamefont
  {Liu}}, \bibinfo {author} {\bibfnamefont {N.}~\bibnamefont {Zhang}}, \bibinfo
  {author} {\bibfnamefont {G.}~\bibnamefont {Qu}}, \bibinfo {author}
  {\bibfnamefont {H.}~\bibnamefont {Ji}}, \bibinfo {author} {\bibfnamefont
  {J.}~\bibnamefont {Han}}, \bibinfo {author} {\bibfnamefont {L.}~\bibnamefont
  {Ge}}, \bibinfo {author} {\bibfnamefont {Y.}~\bibnamefont {Kivshar}},\ and\
  \bibinfo {author} {\bibfnamefont {Q.}~\bibnamefont {Song}},\ }\bibfield
  {title} {\bibinfo {title} {Ultrafast control of vortex microlasers},\ }\href
  {https://doi.org/10.1126/science.aba4597} {\bibfield  {journal} {\bibinfo
  {journal} {Science}\ }\textbf {\bibinfo {volume} {367}},\ \bibinfo {pages}
  {1018} (\bibinfo {year} {2020})}\BibitemShut {NoStop}%
\bibitem [{\citenamefont {Ardizzone}\ \emph {et~al.}(2022)\citenamefont
  {Ardizzone}, \citenamefont {Riminucci}, \citenamefont {Zanotti},
  \citenamefont {Gianfrate}, \citenamefont {{Efthymiou-Tsironi}}, \citenamefont
  {{Su{\`a}rez-Forero}}, \citenamefont {Todisco}, \citenamefont {De~Giorgi},
  \citenamefont {Trypogeorgos}, \citenamefont {Gigli}, \citenamefont {Baldwin},
  \citenamefont {Pfeiffer}, \citenamefont {Ballarini}, \citenamefont {Nguyen},
  \citenamefont {Gerace},\ and\ \citenamefont
  {Sanvitto}}]{Ardizzone_Polariton_2022}%
  \BibitemOpen
  \bibfield  {author} {\bibinfo {author} {\bibfnamefont {V.}~\bibnamefont
  {Ardizzone}}, \bibinfo {author} {\bibfnamefont {F.}~\bibnamefont
  {Riminucci}}, \bibinfo {author} {\bibfnamefont {S.}~\bibnamefont {Zanotti}},
  \bibinfo {author} {\bibfnamefont {A.}~\bibnamefont {Gianfrate}}, \bibinfo
  {author} {\bibfnamefont {M.}~\bibnamefont {{Efthymiou-Tsironi}}}, \bibinfo
  {author} {\bibfnamefont {D.~G.}\ \bibnamefont {{Su{\`a}rez-Forero}}},
  \bibinfo {author} {\bibfnamefont {F.}~\bibnamefont {Todisco}}, \bibinfo
  {author} {\bibfnamefont {M.}~\bibnamefont {De~Giorgi}}, \bibinfo {author}
  {\bibfnamefont {D.}~\bibnamefont {Trypogeorgos}}, \bibinfo {author}
  {\bibfnamefont {G.}~\bibnamefont {Gigli}}, \bibinfo {author} {\bibfnamefont
  {K.}~\bibnamefont {Baldwin}}, \bibinfo {author} {\bibfnamefont
  {L.}~\bibnamefont {Pfeiffer}}, \bibinfo {author} {\bibfnamefont
  {D.}~\bibnamefont {Ballarini}}, \bibinfo {author} {\bibfnamefont {H.~S.}\
  \bibnamefont {Nguyen}}, \bibinfo {author} {\bibfnamefont {D.}~\bibnamefont
  {Gerace}},\ and\ \bibinfo {author} {\bibfnamefont {D.}~\bibnamefont
  {Sanvitto}},\ }\bibfield  {title} {\bibinfo {title} {Polariton
  {{Bose}}--{{Einstein}} condensate from a bound state in the continuum},\
  }\href {https://doi.org/10.1038/s41586-022-04583-7} {\bibfield  {journal}
  {\bibinfo  {journal} {Nature}\ }\textbf {\bibinfo {volume} {605}},\ \bibinfo
  {pages} {447} (\bibinfo {year} {2022})}\BibitemShut {NoStop}%
\bibitem [{\citenamefont {Wu}\ \emph {et~al.}(2024)\citenamefont {Wu},
  \citenamefont {Zhang}, \citenamefont {Song}, \citenamefont {Deng},
  \citenamefont {Du}, \citenamefont {Zeng}, \citenamefont {Zhang},
  \citenamefont {Zhang}, \citenamefont {Chen}, \citenamefont {Wang},
  \citenamefont {Jiang}, \citenamefont {Zhong}, \citenamefont {Wu},
  \citenamefont {Zhu}, \citenamefont {Liang}, \citenamefont {Zhang},
  \citenamefont {Xiong},\ and\ \citenamefont {Liu}}]{Wu_Exciton_2024}%
  \BibitemOpen
  \bibfield  {author} {\bibinfo {author} {\bibfnamefont {X.}~\bibnamefont
  {Wu}}, \bibinfo {author} {\bibfnamefont {S.}~\bibnamefont {Zhang}}, \bibinfo
  {author} {\bibfnamefont {J.}~\bibnamefont {Song}}, \bibinfo {author}
  {\bibfnamefont {X.}~\bibnamefont {Deng}}, \bibinfo {author} {\bibfnamefont
  {W.}~\bibnamefont {Du}}, \bibinfo {author} {\bibfnamefont {X.}~\bibnamefont
  {Zeng}}, \bibinfo {author} {\bibfnamefont {Y.}~\bibnamefont {Zhang}},
  \bibinfo {author} {\bibfnamefont {Z.}~\bibnamefont {Zhang}}, \bibinfo
  {author} {\bibfnamefont {Y.}~\bibnamefont {Chen}}, \bibinfo {author}
  {\bibfnamefont {Y.}~\bibnamefont {Wang}}, \bibinfo {author} {\bibfnamefont
  {C.}~\bibnamefont {Jiang}}, \bibinfo {author} {\bibfnamefont
  {Y.}~\bibnamefont {Zhong}}, \bibinfo {author} {\bibfnamefont
  {B.}~\bibnamefont {Wu}}, \bibinfo {author} {\bibfnamefont {Z.}~\bibnamefont
  {Zhu}}, \bibinfo {author} {\bibfnamefont {Y.}~\bibnamefont {Liang}}, \bibinfo
  {author} {\bibfnamefont {Q.}~\bibnamefont {Zhang}}, \bibinfo {author}
  {\bibfnamefont {Q.}~\bibnamefont {Xiong}},\ and\ \bibinfo {author}
  {\bibfnamefont {X.}~\bibnamefont {Liu}},\ }\bibfield  {title} {\bibinfo
  {title} {Exciton polariton condensation from bound states in the continuum at
  room temperature},\ }\href {https://doi.org/10.1038/s41467-024-47669-8}
  {\bibfield  {journal} {\bibinfo  {journal} {Nat. Commun.}\ }\textbf {\bibinfo
  {volume} {15}},\ \bibinfo {pages} {3345} (\bibinfo {year}
  {2024})}\BibitemShut {NoStop}%
\bibitem [{\citenamefont {Liu}\ \emph {et~al.}(2019{\natexlab{a}})\citenamefont
  {Liu}, \citenamefont {Wang}, \citenamefont {Zhang}, \citenamefont {Wang},
  \citenamefont {Zhao}, \citenamefont {Guan}, \citenamefont {Liu},
  \citenamefont {Shi},\ and\ \citenamefont {Zi}}]{Liu_Circularly_2019}%
  \BibitemOpen
  \bibfield  {author} {\bibinfo {author} {\bibfnamefont {W.}~\bibnamefont
  {Liu}}, \bibinfo {author} {\bibfnamefont {B.}~\bibnamefont {Wang}}, \bibinfo
  {author} {\bibfnamefont {Y.}~\bibnamefont {Zhang}}, \bibinfo {author}
  {\bibfnamefont {J.}~\bibnamefont {Wang}}, \bibinfo {author} {\bibfnamefont
  {M.}~\bibnamefont {Zhao}}, \bibinfo {author} {\bibfnamefont {F.}~\bibnamefont
  {Guan}}, \bibinfo {author} {\bibfnamefont {X.}~\bibnamefont {Liu}}, \bibinfo
  {author} {\bibfnamefont {L.}~\bibnamefont {Shi}},\ and\ \bibinfo {author}
  {\bibfnamefont {J.}~\bibnamefont {Zi}},\ }\bibfield  {title} {\bibinfo
  {title} {Circularly {{Polarized States Spawning}} from {{Bound States}} in
  the {{Continuum}}},\ }\href {https://doi.org/10.1103/PhysRevLett.123.116104}
  {\bibfield  {journal} {\bibinfo  {journal} {Phys. Rev. Lett.}\ }\textbf
  {\bibinfo {volume} {123}},\ \bibinfo {pages} {116104} (\bibinfo {year}
  {2019}{\natexlab{a}})}\BibitemShut {NoStop}%
\bibitem [{\citenamefont {Chen}\ \emph {et~al.}(2023)\citenamefont {Chen},
  \citenamefont {Deng}, \citenamefont {Sha}, \citenamefont {Chen},
  \citenamefont {Wang}, \citenamefont {Chen}, \citenamefont {Wu}, \citenamefont
  {Chu}, \citenamefont {Kivshar}, \citenamefont {Xiao},\ and\ \citenamefont
  {Qiu}}]{Chen_Observation_2023a}%
  \BibitemOpen
  \bibfield  {author} {\bibinfo {author} {\bibfnamefont {Y.}~\bibnamefont
  {Chen}}, \bibinfo {author} {\bibfnamefont {H.}~\bibnamefont {Deng}}, \bibinfo
  {author} {\bibfnamefont {X.}~\bibnamefont {Sha}}, \bibinfo {author}
  {\bibfnamefont {W.}~\bibnamefont {Chen}}, \bibinfo {author} {\bibfnamefont
  {R.}~\bibnamefont {Wang}}, \bibinfo {author} {\bibfnamefont {Y.-H.}\
  \bibnamefont {Chen}}, \bibinfo {author} {\bibfnamefont {D.}~\bibnamefont
  {Wu}}, \bibinfo {author} {\bibfnamefont {J.}~\bibnamefont {Chu}}, \bibinfo
  {author} {\bibfnamefont {Y.~S.}\ \bibnamefont {Kivshar}}, \bibinfo {author}
  {\bibfnamefont {S.}~\bibnamefont {Xiao}},\ and\ \bibinfo {author}
  {\bibfnamefont {C.-W.}\ \bibnamefont {Qiu}},\ }\bibfield  {title} {\bibinfo
  {title} {Observation of intrinsic chiral bound states in the continuum},\
  }\href {https://doi.org/10.1038/s41586-022-05467-6} {\bibfield  {journal}
  {\bibinfo  {journal} {Nature}\ }\textbf {\bibinfo {volume} {613}},\ \bibinfo
  {pages} {474} (\bibinfo {year} {2023})}\BibitemShut {NoStop}%
\bibitem [{\citenamefont {Kodigala}\ \emph {et~al.}(2017)\citenamefont
  {Kodigala}, \citenamefont {Lepetit}, \citenamefont {Gu}, \citenamefont
  {Bahari}, \citenamefont {Fainman},\ and\ \citenamefont
  {Kant{\'e}}}]{Kodigala_Lasing_2017}%
  \BibitemOpen
  \bibfield  {author} {\bibinfo {author} {\bibfnamefont {A.}~\bibnamefont
  {Kodigala}}, \bibinfo {author} {\bibfnamefont {T.}~\bibnamefont {Lepetit}},
  \bibinfo {author} {\bibfnamefont {Q.}~\bibnamefont {Gu}}, \bibinfo {author}
  {\bibfnamefont {B.}~\bibnamefont {Bahari}}, \bibinfo {author} {\bibfnamefont
  {Y.}~\bibnamefont {Fainman}},\ and\ \bibinfo {author} {\bibfnamefont
  {B.}~\bibnamefont {Kant{\'e}}},\ }\bibfield  {title} {\bibinfo {title}
  {Lasing action from photonic bound states in continuum},\ }\href
  {https://doi.org/10.1038/nature20799} {\bibfield  {journal} {\bibinfo
  {journal} {Nature}\ }\textbf {\bibinfo {volume} {541}},\ \bibinfo {pages}
  {196} (\bibinfo {year} {2017})}\BibitemShut {NoStop}%
\bibitem [{\citenamefont {Hwang}\ \emph {et~al.}(2021)\citenamefont {Hwang},
  \citenamefont {Lee}, \citenamefont {Kim}, \citenamefont {Jeong},
  \citenamefont {Kwon}, \citenamefont {Koshelev}, \citenamefont {Kivshar},\
  and\ \citenamefont {Park}}]{Hwang_Ultralowthreshold_2021}%
  \BibitemOpen
  \bibfield  {author} {\bibinfo {author} {\bibfnamefont {M.-S.}\ \bibnamefont
  {Hwang}}, \bibinfo {author} {\bibfnamefont {H.-C.}\ \bibnamefont {Lee}},
  \bibinfo {author} {\bibfnamefont {K.-H.}\ \bibnamefont {Kim}}, \bibinfo
  {author} {\bibfnamefont {K.-Y.}\ \bibnamefont {Jeong}}, \bibinfo {author}
  {\bibfnamefont {S.-H.}\ \bibnamefont {Kwon}}, \bibinfo {author}
  {\bibfnamefont {K.}~\bibnamefont {Koshelev}}, \bibinfo {author}
  {\bibfnamefont {Y.}~\bibnamefont {Kivshar}},\ and\ \bibinfo {author}
  {\bibfnamefont {H.-G.}\ \bibnamefont {Park}},\ }\bibfield  {title} {\bibinfo
  {title} {Ultralow-threshold laser using super-bound states in the
  continuum},\ }\href {https://doi.org/10.1038/s41467-021-24502-0} {\bibfield
  {journal} {\bibinfo  {journal} {Nat. Commun.}\ }\textbf {\bibinfo {volume}
  {12}},\ \bibinfo {pages} {4135} (\bibinfo {year} {2021})}\BibitemShut
  {NoStop}%
\bibitem [{\citenamefont {Yu}\ \emph {et~al.}(2021)\citenamefont {Yu},
  \citenamefont {Sakanas}, \citenamefont {Zali}, \citenamefont {Semenova},
  \citenamefont {Yvind},\ and\ \citenamefont
  {M{\o}rk}}]{Yu_Ultracoherent_2021}%
  \BibitemOpen
  \bibfield  {author} {\bibinfo {author} {\bibfnamefont {Y.}~\bibnamefont
  {Yu}}, \bibinfo {author} {\bibfnamefont {A.}~\bibnamefont {Sakanas}},
  \bibinfo {author} {\bibfnamefont {A.~R.}\ \bibnamefont {Zali}}, \bibinfo
  {author} {\bibfnamefont {E.}~\bibnamefont {Semenova}}, \bibinfo {author}
  {\bibfnamefont {K.}~\bibnamefont {Yvind}},\ and\ \bibinfo {author}
  {\bibfnamefont {J.}~\bibnamefont {M{\o}rk}},\ }\bibfield  {title} {\bibinfo
  {title} {Ultra-coherent {{Fano}} laser based on a bound state in the
  continuum},\ }\href {https://doi.org/10.1038/s41566-021-00860-5} {\bibfield
  {journal} {\bibinfo  {journal} {Nat. Photonics}\ }\textbf {\bibinfo {volume}
  {15}},\ \bibinfo {pages} {758} (\bibinfo {year} {2021})}\BibitemShut
  {NoStop}%
\bibitem [{\citenamefont {Ren}\ \emph {et~al.}(2022)\citenamefont {Ren},
  \citenamefont {Li}, \citenamefont {Liu}, \citenamefont {Chen}, \citenamefont
  {Chen}, \citenamefont {Peng},\ and\ \citenamefont
  {Liu}}]{Ren_Lowthreshold_2022}%
  \BibitemOpen
  \bibfield  {author} {\bibinfo {author} {\bibfnamefont {Y.}~\bibnamefont
  {Ren}}, \bibinfo {author} {\bibfnamefont {P.}~\bibnamefont {Li}}, \bibinfo
  {author} {\bibfnamefont {Z.}~\bibnamefont {Liu}}, \bibinfo {author}
  {\bibfnamefont {Z.}~\bibnamefont {Chen}}, \bibinfo {author} {\bibfnamefont
  {Y.-L.}\ \bibnamefont {Chen}}, \bibinfo {author} {\bibfnamefont
  {C.}~\bibnamefont {Peng}},\ and\ \bibinfo {author} {\bibfnamefont
  {J.}~\bibnamefont {Liu}},\ }\bibfield  {title} {\bibinfo {title}
  {Low-threshold nanolasers based on miniaturized bound states in the
  continuum},\ }\href {https://doi.org/10.1126/sciadv.ade8817} {\bibfield
  {journal} {\bibinfo  {journal} {Sci. Adv.}\ }\textbf {\bibinfo {volume}
  {8}},\ \bibinfo {pages} {eade8817} (\bibinfo {year} {2022})}\BibitemShut
  {NoStop}%
\bibitem [{\citenamefont {Liu}\ \emph {et~al.}(2019{\natexlab{b}})\citenamefont
  {Liu}, \citenamefont {Xu}, \citenamefont {Lin}, \citenamefont {Xiang},
  \citenamefont {Feng}, \citenamefont {Cao}, \citenamefont {Li}, \citenamefont
  {Lan},\ and\ \citenamefont {Liu}}]{Liu_High_2019}%
  \BibitemOpen
  \bibfield  {author} {\bibinfo {author} {\bibfnamefont {Z.}~\bibnamefont
  {Liu}}, \bibinfo {author} {\bibfnamefont {Y.}~\bibnamefont {Xu}}, \bibinfo
  {author} {\bibfnamefont {Y.}~\bibnamefont {Lin}}, \bibinfo {author}
  {\bibfnamefont {J.}~\bibnamefont {Xiang}}, \bibinfo {author} {\bibfnamefont
  {T.}~\bibnamefont {Feng}}, \bibinfo {author} {\bibfnamefont {Q.}~\bibnamefont
  {Cao}}, \bibinfo {author} {\bibfnamefont {J.}~\bibnamefont {Li}}, \bibinfo
  {author} {\bibfnamefont {S.}~\bibnamefont {Lan}},\ and\ \bibinfo {author}
  {\bibfnamefont {J.}~\bibnamefont {Liu}},\ }\bibfield  {title} {\bibinfo
  {title} {High-{{Q}} {{Quasibound States}} in the {{Continuum}} for
  {{Nonlinear Metasurfaces}}},\ }\href
  {https://doi.org/10.1103/PhysRevLett.123.253901} {\bibfield  {journal}
  {\bibinfo  {journal} {Phys. Rev. Lett.}\ }\textbf {\bibinfo {volume} {123}},\
  \bibinfo {pages} {253901} (\bibinfo {year} {2019}{\natexlab{b}})}\BibitemShut
  {NoStop}%
\bibitem [{\citenamefont {Koshelev}\ \emph {et~al.}(2020)\citenamefont
  {Koshelev}, \citenamefont {Kruk}, \citenamefont {{Melik-Gaykazyan}},
  \citenamefont {Choi}, \citenamefont {Bogdanov}, \citenamefont {Park},\ and\
  \citenamefont {Kivshar}}]{Koshelev_Subwavelength_2020}%
  \BibitemOpen
  \bibfield  {author} {\bibinfo {author} {\bibfnamefont {K.}~\bibnamefont
  {Koshelev}}, \bibinfo {author} {\bibfnamefont {S.}~\bibnamefont {Kruk}},
  \bibinfo {author} {\bibfnamefont {E.}~\bibnamefont {{Melik-Gaykazyan}}},
  \bibinfo {author} {\bibfnamefont {J.-H.}\ \bibnamefont {Choi}}, \bibinfo
  {author} {\bibfnamefont {A.}~\bibnamefont {Bogdanov}}, \bibinfo {author}
  {\bibfnamefont {H.-G.}\ \bibnamefont {Park}},\ and\ \bibinfo {author}
  {\bibfnamefont {Y.}~\bibnamefont {Kivshar}},\ }\bibfield  {title} {\bibinfo
  {title} {Subwavelength dielectric resonators for nonlinear nanophotonics},\
  }\href {https://doi.org/10.1126/science.aaz3985} {\bibfield  {journal}
  {\bibinfo  {journal} {Science}\ }\textbf {\bibinfo {volume} {367}},\ \bibinfo
  {pages} {288} (\bibinfo {year} {2020})}\BibitemShut {NoStop}%
\bibitem [{\citenamefont {Schiattarella}\ \emph {et~al.}(2024)\citenamefont
  {Schiattarella}, \citenamefont {Romano}, \citenamefont {Sirleto},
  \citenamefont {Mocella}, \citenamefont {Rendina}, \citenamefont {Lanzio},
  \citenamefont {Riminucci}, \citenamefont {Schwartzberg}, \citenamefont
  {Cabrini}, \citenamefont {Chen}, \citenamefont {Liang}, \citenamefont {Liu},\
  and\ \citenamefont {Zito}}]{Schiattarella_Directive_2024}%
  \BibitemOpen
  \bibfield  {author} {\bibinfo {author} {\bibfnamefont {C.}~\bibnamefont
  {Schiattarella}}, \bibinfo {author} {\bibfnamefont {S.}~\bibnamefont
  {Romano}}, \bibinfo {author} {\bibfnamefont {L.}~\bibnamefont {Sirleto}},
  \bibinfo {author} {\bibfnamefont {V.}~\bibnamefont {Mocella}}, \bibinfo
  {author} {\bibfnamefont {I.}~\bibnamefont {Rendina}}, \bibinfo {author}
  {\bibfnamefont {V.}~\bibnamefont {Lanzio}}, \bibinfo {author} {\bibfnamefont
  {F.}~\bibnamefont {Riminucci}}, \bibinfo {author} {\bibfnamefont
  {A.}~\bibnamefont {Schwartzberg}}, \bibinfo {author} {\bibfnamefont
  {S.}~\bibnamefont {Cabrini}}, \bibinfo {author} {\bibfnamefont
  {J.}~\bibnamefont {Chen}}, \bibinfo {author} {\bibfnamefont {L.}~\bibnamefont
  {Liang}}, \bibinfo {author} {\bibfnamefont {X.}~\bibnamefont {Liu}},\ and\
  \bibinfo {author} {\bibfnamefont {G.}~\bibnamefont {Zito}},\ }\bibfield
  {title} {\bibinfo {title} {Directive giant upconversion by supercritical
  bound states in the continuum},\ }\href
  {https://doi.org/10.1038/s41586-023-06967-9} {\bibfield  {journal} {\bibinfo
  {journal} {Nature}\ }\textbf {\bibinfo {volume} {626}},\ \bibinfo {pages}
  {765} (\bibinfo {year} {2024})}\BibitemShut {NoStop}%
\bibitem [{\citenamefont {Liu}\ \emph {et~al.}(2023)\citenamefont {Liu},
  \citenamefont {Guo}, \citenamefont {Tan}, \citenamefont {Hu}, \citenamefont
  {Sun}, \citenamefont {Fan}, \citenamefont {Zhang}, \citenamefont {Jin},\ and\
  \citenamefont {He}}]{Liu_Phase_2023}%
  \BibitemOpen
  \bibfield  {author} {\bibinfo {author} {\bibfnamefont {Z.}~\bibnamefont
  {Liu}}, \bibinfo {author} {\bibfnamefont {T.}~\bibnamefont {Guo}}, \bibinfo
  {author} {\bibfnamefont {Q.}~\bibnamefont {Tan}}, \bibinfo {author}
  {\bibfnamefont {Z.}~\bibnamefont {Hu}}, \bibinfo {author} {\bibfnamefont
  {Y.}~\bibnamefont {Sun}}, \bibinfo {author} {\bibfnamefont {H.}~\bibnamefont
  {Fan}}, \bibinfo {author} {\bibfnamefont {Z.}~\bibnamefont {Zhang}}, \bibinfo
  {author} {\bibfnamefont {Y.}~\bibnamefont {Jin}},\ and\ \bibinfo {author}
  {\bibfnamefont {S.}~\bibnamefont {He}},\ }\bibfield  {title} {\bibinfo
  {title} {Phase {{Interrogation Sensor Based}} on {{All-Dielectric BIC
  Metasurface}}},\ }\href {https://doi.org/10.1021/acs.nanolett.3c03089}
  {\bibfield  {journal} {\bibinfo  {journal} {Nano Lett.}\ }\textbf {\bibinfo
  {volume} {23}},\ \bibinfo {pages} {10441} (\bibinfo {year}
  {2023})}\BibitemShut {NoStop}%
\bibitem [{\citenamefont {Skirlo}\ \emph {et~al.}(2014)\citenamefont {Skirlo},
  \citenamefont {Lu},\ and\ \citenamefont {Solja{\v
  c}i{\'c}}}]{Skirlo_Multimode_2014}%
  \BibitemOpen
  \bibfield  {author} {\bibinfo {author} {\bibfnamefont {S.~A.}\ \bibnamefont
  {Skirlo}}, \bibinfo {author} {\bibfnamefont {L.}~\bibnamefont {Lu}},\ and\
  \bibinfo {author} {\bibfnamefont {M.}~\bibnamefont {Solja{\v c}i{\'c}}},\
  }\bibfield  {title} {\bibinfo {title} {Multimode {{One-Way Waveguides}} of
  {{Large Chern Numbers}}},\ }\href
  {https://doi.org/10.1103/PhysRevLett.113.113904} {\bibfield  {journal}
  {\bibinfo  {journal} {Phys. Rev. Lett.}\ }\textbf {\bibinfo {volume} {113}},\
  \bibinfo {pages} {113904} (\bibinfo {year} {2014})}\BibitemShut {NoStop}%
\bibitem [{\citenamefont {Lu}\ \emph {et~al.}(2016)\citenamefont {Lu},
  \citenamefont {Joannopoulos},\ and\ \citenamefont {Solja{\v
  c}i{\'c}}}]{Lu_Topological_2016}%
  \BibitemOpen
  \bibfield  {author} {\bibinfo {author} {\bibfnamefont {L.}~\bibnamefont
  {Lu}}, \bibinfo {author} {\bibfnamefont {J.~D.}\ \bibnamefont
  {Joannopoulos}},\ and\ \bibinfo {author} {\bibfnamefont {M.}~\bibnamefont
  {Solja{\v c}i{\'c}}},\ }\bibfield  {title} {\bibinfo {title} {Topological
  states in photonic systems},\ }\href {https://doi.org/10.1038/nphys3796}
  {\bibfield  {journal} {\bibinfo  {journal} {Nat. Phys.}\ }\textbf {\bibinfo
  {volume} {12}},\ \bibinfo {pages} {626} (\bibinfo {year} {2016})}\BibitemShut
  {NoStop}%
\bibitem [{\citenamefont {Wu}\ \emph {et~al.}(2017)\citenamefont {Wu},
  \citenamefont {Li}, \citenamefont {Hu}, \citenamefont {Ao}, \citenamefont
  {Zhao},\ and\ \citenamefont {Gong}}]{Wu_Applications_2017}%
  \BibitemOpen
  \bibfield  {author} {\bibinfo {author} {\bibfnamefont {Y.}~\bibnamefont
  {Wu}}, \bibinfo {author} {\bibfnamefont {C.}~\bibnamefont {Li}}, \bibinfo
  {author} {\bibfnamefont {X.}~\bibnamefont {Hu}}, \bibinfo {author}
  {\bibfnamefont {Y.}~\bibnamefont {Ao}}, \bibinfo {author} {\bibfnamefont
  {Y.}~\bibnamefont {Zhao}},\ and\ \bibinfo {author} {\bibfnamefont
  {Q.}~\bibnamefont {Gong}},\ }\bibfield  {title} {\bibinfo {title}
  {Applications of {{Topological Photonics}} in {{Integrated Photonic
  Devices}}},\ }\href {https://doi.org/10.1002/adom.201700357} {\bibfield
  {journal} {\bibinfo  {journal} {Adv. Opt. Mater.}\ }\textbf {\bibinfo
  {volume} {5}},\ \bibinfo {pages} {1700357} (\bibinfo {year}
  {2017})}\BibitemShut {NoStop}%
\bibitem [{\citenamefont {Chong}\ \emph {et~al.}(2008)\citenamefont {Chong},
  \citenamefont {Wen},\ and\ \citenamefont {Solja{\v
  c}i{\'c}}}]{Chong_Effective_2008}%
  \BibitemOpen
  \bibfield  {author} {\bibinfo {author} {\bibfnamefont {Y.~D.}\ \bibnamefont
  {Chong}}, \bibinfo {author} {\bibfnamefont {X.-G.}\ \bibnamefont {Wen}},\
  and\ \bibinfo {author} {\bibfnamefont {M.}~\bibnamefont {Solja{\v
  c}i{\'c}}},\ }\bibfield  {title} {\bibinfo {title} {Effective theory of
  quadratic degeneracies},\ }\href {https://doi.org/10.1103/PhysRevB.77.235125}
  {\bibfield  {journal} {\bibinfo  {journal} {Phys. Rev. B}\ }\textbf {\bibinfo
  {volume} {77}},\ \bibinfo {pages} {235125} (\bibinfo {year}
  {2008})}\BibitemShut {NoStop}%
\bibitem [{\citenamefont {Mei}\ \emph {et~al.}(2012)\citenamefont {Mei},
  \citenamefont {Wu}, \citenamefont {Chan},\ and\ \citenamefont
  {Zhang}}]{Mei_Firstprinciples_2012}%
  \BibitemOpen
  \bibfield  {author} {\bibinfo {author} {\bibfnamefont {J.}~\bibnamefont
  {Mei}}, \bibinfo {author} {\bibfnamefont {Y.}~\bibnamefont {Wu}}, \bibinfo
  {author} {\bibfnamefont {C.~T.}\ \bibnamefont {Chan}},\ and\ \bibinfo
  {author} {\bibfnamefont {Z.-Q.}\ \bibnamefont {Zhang}},\ }\bibfield  {title}
  {\bibinfo {title} {First-principles study of {{Dirac}} and {{Dirac-like}}
  cones in phononic and photonic crystals},\ }\href
  {https://doi.org/10.1103/PhysRevB.86.035141} {\bibfield  {journal} {\bibinfo
  {journal} {Phys. Rev. B}\ }\textbf {\bibinfo {volume} {86}},\ \bibinfo
  {pages} {035141} (\bibinfo {year} {2012})}\BibitemShut {NoStop}%
\bibitem [{\citenamefont {Gao}\ \emph {et~al.}(2023)\citenamefont {Gao},
  \citenamefont {Zhao}, \citenamefont {Wu}, \citenamefont {Feng}, \citenamefont
  {Zhang}, \citenamefont {Qiao}, \citenamefont {Chiu},\ and\ \citenamefont
  {Feng}}]{Gao_Topological_2023a}%
  \BibitemOpen
  \bibfield  {author} {\bibinfo {author} {\bibfnamefont {Z.}~\bibnamefont
  {Gao}}, \bibinfo {author} {\bibfnamefont {H.}~\bibnamefont {Zhao}}, \bibinfo
  {author} {\bibfnamefont {T.}~\bibnamefont {Wu}}, \bibinfo {author}
  {\bibfnamefont {X.}~\bibnamefont {Feng}}, \bibinfo {author} {\bibfnamefont
  {Z.}~\bibnamefont {Zhang}}, \bibinfo {author} {\bibfnamefont
  {X.}~\bibnamefont {Qiao}}, \bibinfo {author} {\bibfnamefont {C.-K.}\
  \bibnamefont {Chiu}},\ and\ \bibinfo {author} {\bibfnamefont
  {L.}~\bibnamefont {Feng}},\ }\bibfield  {title} {\bibinfo {title}
  {Topological quadratic-node semimetal in a photonic microring lattice},\
  }\href {https://doi.org/10.1038/s41467-023-38861-3} {\bibfield  {journal}
  {\bibinfo  {journal} {Nat Commun.}\ }\textbf {\bibinfo {volume} {14}},\
  \bibinfo {pages} {3206} (\bibinfo {year} {2023})}\BibitemShut {NoStop}%
\bibitem [{\citenamefont {Ochiai}\ and\ \citenamefont
  {Onoda}(2009)}]{Ochiai_Photonic_2009}%
  \BibitemOpen
  \bibfield  {author} {\bibinfo {author} {\bibfnamefont {T.}~\bibnamefont
  {Ochiai}}\ and\ \bibinfo {author} {\bibfnamefont {M.}~\bibnamefont {Onoda}},\
  }\bibfield  {title} {\bibinfo {title} {Photonic analog of graphene model and
  its extension: {{Dirac}} cone, symmetry, and edge states},\ }\href
  {https://doi.org/10.1103/PhysRevB.80.155103} {\bibfield  {journal} {\bibinfo
  {journal} {Phys. Rev. B}\ }\textbf {\bibinfo {volume} {80}},\ \bibinfo
  {pages} {155103} (\bibinfo {year} {2009})}\BibitemShut {NoStop}%
\bibitem [{\citenamefont {Collins}\ \emph {et~al.}(2016)\citenamefont
  {Collins}, \citenamefont {Zhang}, \citenamefont {Bojko}, \citenamefont
  {Chrostowski},\ and\ \citenamefont {Rechtsman}}]{Collins_Integrated_2016}%
  \BibitemOpen
  \bibfield  {author} {\bibinfo {author} {\bibfnamefont {M.~J.}\ \bibnamefont
  {Collins}}, \bibinfo {author} {\bibfnamefont {F.}~\bibnamefont {Zhang}},
  \bibinfo {author} {\bibfnamefont {R.}~\bibnamefont {Bojko}}, \bibinfo
  {author} {\bibfnamefont {L.}~\bibnamefont {Chrostowski}},\ and\ \bibinfo
  {author} {\bibfnamefont {M.~C.}\ \bibnamefont {Rechtsman}},\ }\bibfield
  {title} {\bibinfo {title} {Integrated optical {{Dirac}} physics via inversion
  symmetry breaking},\ }\href {https://doi.org/10.1103/PhysRevA.94.063827}
  {\bibfield  {journal} {\bibinfo  {journal} {Phys. Rev. A}\ }\textbf {\bibinfo
  {volume} {94}},\ \bibinfo {pages} {063827} (\bibinfo {year}
  {2016})}\BibitemShut {NoStop}%
\bibitem [{\citenamefont {Zhang}\ \emph
  {et~al.}(2018{\natexlab{b}})\citenamefont {Zhang}, \citenamefont {Gogna},
  \citenamefont {Burg}, \citenamefont {Tutuc},\ and\ \citenamefont
  {Deng}}]{Zhang_Photoniccrystal_2018a}%
  \BibitemOpen
  \bibfield  {author} {\bibinfo {author} {\bibfnamefont {L.}~\bibnamefont
  {Zhang}}, \bibinfo {author} {\bibfnamefont {R.}~\bibnamefont {Gogna}},
  \bibinfo {author} {\bibfnamefont {W.}~\bibnamefont {Burg}}, \bibinfo {author}
  {\bibfnamefont {E.}~\bibnamefont {Tutuc}},\ and\ \bibinfo {author}
  {\bibfnamefont {H.}~\bibnamefont {Deng}},\ }\bibfield  {title} {\bibinfo
  {title} {Photonic-crystal exciton-polaritons in monolayer semiconductors},\
  }\href {https://doi.org/10.1038/s41467-018-03188-x} {\bibfield  {journal}
  {\bibinfo  {journal} {Nat. Commun.}\ }\textbf {\bibinfo {volume} {9}},\
  \bibinfo {pages} {713} (\bibinfo {year} {2018}{\natexlab{b}})}\BibitemShut
  {NoStop}%
\bibitem [{\citenamefont {Cunningham}\ \emph {et~al.}(2019)\citenamefont
  {Cunningham}, \citenamefont {Hanbicki}, \citenamefont {Reinecke},
  \citenamefont {McCreary},\ and\ \citenamefont
  {Jonker}}]{Cunningham_Resonant_2019}%
  \BibitemOpen
  \bibfield  {author} {\bibinfo {author} {\bibfnamefont {P.~D.}\ \bibnamefont
  {Cunningham}}, \bibinfo {author} {\bibfnamefont {A.~T.}\ \bibnamefont
  {Hanbicki}}, \bibinfo {author} {\bibfnamefont {T.~L.}\ \bibnamefont
  {Reinecke}}, \bibinfo {author} {\bibfnamefont {K.~M.}\ \bibnamefont
  {McCreary}},\ and\ \bibinfo {author} {\bibfnamefont {B.~T.}\ \bibnamefont
  {Jonker}},\ }\bibfield  {title} {\bibinfo {title} {Resonant optical {{Stark}}
  effect in monolayer {{WS}}$_2$},\ }\href
  {https://doi.org/10.1038/s41467-019-13501-x} {\bibfield  {journal} {\bibinfo
  {journal} {Nat. Commun.}\ }\textbf {\bibinfo {volume} {10}},\ \bibinfo
  {pages} {5539} (\bibinfo {year} {2019})}\BibitemShut {NoStop}%
\bibitem [{\citenamefont {Yong}\ \emph {et~al.}(2019)\citenamefont {Yong},
  \citenamefont {Utama}, \citenamefont {Ong}, \citenamefont {Cao},
  \citenamefont {Regan}, \citenamefont {Horng}, \citenamefont {Shen},
  \citenamefont {Cai}, \citenamefont {Watanabe}, \citenamefont {Taniguchi},
  \citenamefont {Tongay}, \citenamefont {Deng}, \citenamefont {Zettl},
  \citenamefont {Louie},\ and\ \citenamefont
  {Wang}}]{Yong_Valleydependent_2019}%
  \BibitemOpen
  \bibfield  {author} {\bibinfo {author} {\bibfnamefont {C.-K.}\ \bibnamefont
  {Yong}}, \bibinfo {author} {\bibfnamefont {M.~I.~B.}\ \bibnamefont {Utama}},
  \bibinfo {author} {\bibfnamefont {C.~S.}\ \bibnamefont {Ong}}, \bibinfo
  {author} {\bibfnamefont {T.}~\bibnamefont {Cao}}, \bibinfo {author}
  {\bibfnamefont {E.~C.}\ \bibnamefont {Regan}}, \bibinfo {author}
  {\bibfnamefont {J.}~\bibnamefont {Horng}}, \bibinfo {author} {\bibfnamefont
  {Y.}~\bibnamefont {Shen}}, \bibinfo {author} {\bibfnamefont {H.}~\bibnamefont
  {Cai}}, \bibinfo {author} {\bibfnamefont {K.}~\bibnamefont {Watanabe}},
  \bibinfo {author} {\bibfnamefont {T.}~\bibnamefont {Taniguchi}}, \bibinfo
  {author} {\bibfnamefont {S.}~\bibnamefont {Tongay}}, \bibinfo {author}
  {\bibfnamefont {H.}~\bibnamefont {Deng}}, \bibinfo {author} {\bibfnamefont
  {A.}~\bibnamefont {Zettl}}, \bibinfo {author} {\bibfnamefont {S.~G.}\
  \bibnamefont {Louie}},\ and\ \bibinfo {author} {\bibfnamefont
  {F.}~\bibnamefont {Wang}},\ }\bibfield  {title} {\bibinfo {title}
  {Valley-dependent exciton fine structure and {{Autler}}--{{Townes}} doublets
  from {{Berry}} phases in monolayer {{MoSe}}$_2$},\ }\href
  {https://doi.org/10.1038/s41563-019-0447-8} {\bibfield  {journal} {\bibinfo
  {journal} {Nat. Mater.}\ }\textbf {\bibinfo {volume} {18}},\ \bibinfo {pages}
  {1065} (\bibinfo {year} {2019})}\BibitemShut {NoStop}%
\bibitem [{\citenamefont {Zhou}\ \emph {et~al.}(2024)\citenamefont {Zhou},
  \citenamefont {Liu}, \citenamefont {Liu}, \citenamefont {Lu}, \citenamefont
  {Li}, \citenamefont {Xie}, \citenamefont {Lydick}, \citenamefont {Hao},
  \citenamefont {Liu}, \citenamefont {Watanabe}, \citenamefont {Taniguchi},
  \citenamefont {Chou}, \citenamefont {Forrest},\ and\ \citenamefont
  {Deng}}]{Zhou_Cavity_2024}%
  \BibitemOpen
  \bibfield  {author} {\bibinfo {author} {\bibfnamefont {L.}~\bibnamefont
  {Zhou}}, \bibinfo {author} {\bibfnamefont {B.}~\bibnamefont {Liu}}, \bibinfo
  {author} {\bibfnamefont {Y.}~\bibnamefont {Liu}}, \bibinfo {author}
  {\bibfnamefont {Y.}~\bibnamefont {Lu}}, \bibinfo {author} {\bibfnamefont
  {Q.}~\bibnamefont {Li}}, \bibinfo {author} {\bibfnamefont {X.}~\bibnamefont
  {Xie}}, \bibinfo {author} {\bibfnamefont {N.}~\bibnamefont {Lydick}},
  \bibinfo {author} {\bibfnamefont {R.}~\bibnamefont {Hao}}, \bibinfo {author}
  {\bibfnamefont {C.}~\bibnamefont {Liu}}, \bibinfo {author} {\bibfnamefont
  {K.}~\bibnamefont {Watanabe}}, \bibinfo {author} {\bibfnamefont
  {T.}~\bibnamefont {Taniguchi}}, \bibinfo {author} {\bibfnamefont {Y.-H.}\
  \bibnamefont {Chou}}, \bibinfo {author} {\bibfnamefont {S.~R.}\ \bibnamefont
  {Forrest}},\ and\ \bibinfo {author} {\bibfnamefont {H.}~\bibnamefont
  {Deng}},\ }\bibfield  {title} {\bibinfo {title} {Cavity {{Floquet}}
  engineering},\ }\href {https://doi.org/10.1038/s41467-024-52014-0} {\bibfield
   {journal} {\bibinfo  {journal} {Nat. Commun.}\ }\textbf {\bibinfo {volume}
  {15}},\ \bibinfo {pages} {7782} (\bibinfo {year} {2024})}\BibitemShut
  {NoStop}%
\bibitem [{\citenamefont {Mak}\ and\ \citenamefont
  {Shan}(2022)}]{Mak_Semiconductor_2022a}%
  \BibitemOpen
  \bibfield  {author} {\bibinfo {author} {\bibfnamefont {K.~F.}\ \bibnamefont
  {Mak}}\ and\ \bibinfo {author} {\bibfnamefont {J.}~\bibnamefont {Shan}},\
  }\bibfield  {title} {\bibinfo {title} {Semiconductor moir{\'e} materials},\
  }\href {https://doi.org/10.1038/s41565-022-01165-6} {\bibfield  {journal}
  {\bibinfo  {journal} {Nat. Nanotechnol.}\ }\textbf {\bibinfo {volume} {17}},\
  \bibinfo {pages} {686} (\bibinfo {year} {2022})}\BibitemShut {NoStop}%
\bibitem [{\citenamefont {Chang}\ \emph {et~al.}(2023)\citenamefont {Chang},
  \citenamefont {Liu},\ and\ \citenamefont
  {MacDonald}}]{Chang_Colloquium_2023}%
  \BibitemOpen
  \bibfield  {author} {\bibinfo {author} {\bibfnamefont {C.-Z.}\ \bibnamefont
  {Chang}}, \bibinfo {author} {\bibfnamefont {C.-X.}\ \bibnamefont {Liu}},\
  and\ \bibinfo {author} {\bibfnamefont {A.~H.}\ \bibnamefont {MacDonald}},\
  }\bibfield  {title} {\bibinfo {title} {Colloquium: {{Quantum}} anomalous
  {{Hall}} effect},\ }\href {https://doi.org/10.1103/RevModPhys.95.011002}
  {\bibfield  {journal} {\bibinfo  {journal} {Rev. Mod. Phys.}\ }\textbf
  {\bibinfo {volume} {95}},\ \bibinfo {pages} {011002} (\bibinfo {year}
  {2023})}\BibitemShut {NoStop}%
\bibitem [{\citenamefont {Zhao}\ \emph {et~al.}(2024)\citenamefont {Zhao},
  \citenamefont {Kang}, \citenamefont {Zhang}, \citenamefont {Kn{\"u}ppel},
  \citenamefont {Tao}, \citenamefont {Li}, \citenamefont {Tschirhart},
  \citenamefont {Redekop}, \citenamefont {Watanabe}, \citenamefont {Taniguchi},
  \citenamefont {Young}, \citenamefont {Shan},\ and\ \citenamefont
  {Mak}}]{Zhao_Realization_2024a}%
  \BibitemOpen
  \bibfield  {author} {\bibinfo {author} {\bibfnamefont {W.}~\bibnamefont
  {Zhao}}, \bibinfo {author} {\bibfnamefont {K.}~\bibnamefont {Kang}}, \bibinfo
  {author} {\bibfnamefont {Y.}~\bibnamefont {Zhang}}, \bibinfo {author}
  {\bibfnamefont {P.}~\bibnamefont {Kn{\"u}ppel}}, \bibinfo {author}
  {\bibfnamefont {Z.}~\bibnamefont {Tao}}, \bibinfo {author} {\bibfnamefont
  {L.}~\bibnamefont {Li}}, \bibinfo {author} {\bibfnamefont {C.~L.}\
  \bibnamefont {Tschirhart}}, \bibinfo {author} {\bibfnamefont
  {E.}~\bibnamefont {Redekop}}, \bibinfo {author} {\bibfnamefont
  {K.}~\bibnamefont {Watanabe}}, \bibinfo {author} {\bibfnamefont
  {T.}~\bibnamefont {Taniguchi}}, \bibinfo {author} {\bibfnamefont {A.~F.}\
  \bibnamefont {Young}}, \bibinfo {author} {\bibfnamefont {J.}~\bibnamefont
  {Shan}},\ and\ \bibinfo {author} {\bibfnamefont {K.~F.}\ \bibnamefont
  {Mak}},\ }\bibfield  {title} {\bibinfo {title} {Realization of the {{Haldane
  Chern}} insulator in a moir{\'e} lattice},\ }\href
  {https://doi.org/10.1038/s41567-023-02284-0} {\bibfield  {journal} {\bibinfo
  {journal} {Nat. Phys.}\ }\textbf {\bibinfo {volume} {20}},\ \bibinfo {pages}
  {275} (\bibinfo {year} {2024})}\BibitemShut {NoStop}%
\bibitem [{\citenamefont {Li}\ \emph {et~al.}(2014)\citenamefont {Li},
  \citenamefont {Chernikov}, \citenamefont {Zhang}, \citenamefont {Rigosi},
  \citenamefont {Hill}, \citenamefont {{van der Zande}}, \citenamefont
  {Chenet}, \citenamefont {Shih}, \citenamefont {Hone},\ and\ \citenamefont
  {Heinz}}]{Li_Measurement_2014}%
  \BibitemOpen
  \bibfield  {author} {\bibinfo {author} {\bibfnamefont {Y.}~\bibnamefont
  {Li}}, \bibinfo {author} {\bibfnamefont {A.}~\bibnamefont {Chernikov}},
  \bibinfo {author} {\bibfnamefont {X.}~\bibnamefont {Zhang}}, \bibinfo
  {author} {\bibfnamefont {A.}~\bibnamefont {Rigosi}}, \bibinfo {author}
  {\bibfnamefont {H.~M.}\ \bibnamefont {Hill}}, \bibinfo {author}
  {\bibfnamefont {A.~M.}\ \bibnamefont {{van der Zande}}}, \bibinfo {author}
  {\bibfnamefont {D.~A.}\ \bibnamefont {Chenet}}, \bibinfo {author}
  {\bibfnamefont {E.-M.}\ \bibnamefont {Shih}}, \bibinfo {author}
  {\bibfnamefont {J.}~\bibnamefont {Hone}},\ and\ \bibinfo {author}
  {\bibfnamefont {T.~F.}\ \bibnamefont {Heinz}},\ }\bibfield  {title} {\bibinfo
  {title} {Measurement of the optical dielectric function of monolayer
  transition-metal dichalcogenides: {{MoS}}$_2$, {{MoSe}}$_2$, {{WS}}$_2$, and
  {{WSe}}$_2$},\ }\href {https://doi.org/10.1103/PhysRevB.90.205422} {\bibfield
   {journal} {\bibinfo  {journal} {Phys. Rev. B}\ }\textbf {\bibinfo {volume}
  {90}},\ \bibinfo {pages} {205422} (\bibinfo {year} {2014})}\BibitemShut
  {NoStop}%
\end{thebibliography}%

\end{document}